\documentclass{article}
\usepackage[utf8]{inputenc}

\usepackage{graphicx}
\usepackage{color}
\usepackage{amsmath,amsthm,amssymb,hyperref}
\usepackage{amsfonts}
\usepackage[normal]{subfigure}
\usepackage{a4wide}
\usepackage{enumitem}
\usepackage{hyperref}
\newcommand{\cS}{\mathcal{S}}

\newcommand{\Ub}{{\bf U}}
\newcommand{\vb}{{\bf v}}
\newcommand{\hv}{\hat{v}}
\newcommand{\hw}{\hat{w}}

\newcommand{\x}{x}

\newcommand{\benoit}[1]{{\color{blue}#1}}
\newcommand \commentout[1] {}
\newtheorem{theorem}{Theorem}[]

\begin{document}
\title{A Hamilton-Jacobi approach to nonlocal kinetic equations}
\author{Nadia Loy \thanks{Department of Mathematical Sciences ``G. L. Lagrange'', Politecnico di Torino, Corso Duca degli Abruzzi 24, 10129, Torino, Italy, (\texttt{nadia.loy@polito.it})} \and Benoit Perthame \thanks{Sorbonne Université, CNRS, Université de Paris, Inria, Laboratoire Jacques-Louis
Lions, F-75005 Paris, France,  (\texttt{benoit.perthame@sorbonne-universite.fr})} 
}\date{\small }

\maketitle

\begin{abstract}
   Highly concentrated patterns have been observed in a spatially heterogeneous, nonlocal, model of BGK type implementing a velocity-jump process.
   We study both a linear and a nonlinear case and describe the concentration profile. In particular, we analyse a hyperbolic (or high frequency) regime that can be interpreted both as a local (microscopic) or as a nonlocal (macroscopic) rescaling. We consider a Hopf-Cole transform and derive a Hamilton-Jacobi equation. The concentrations are then explained as a consequence of the stationary points of the Hamiltonian that is spatially heterogeneous like the velocity-jump process. After revising the classical hydrodynamic limits for the aggregate quantities and the eikonal equation that can be derived from those with a Hopf-Cole transform, we find that the Hamilton-Jacobi equation is a second order approximation of the eikonal equation in the limit of small diffusivity. For nonlinear turning kernels, the Hopf-Cole transform allows to study the stability of the possible homogeneous configurations and of patterns and the results of a linear stability analysis previously obtained are found and extended to a nonlinear regime. In particular, it is shown that instability (pattern formation) occurs when the Hamiltonian
is convex-concave.
\end{abstract}

\section*{Introduction}
Kinetic equations have proved to be an effective mathematical framework for modeling cell migration, both for bacteria  \cite{Alt.88,Calvez2015KRM,Othmer_Hillen.00,Othmer_Hillen.02, Chalub_Markowich_Perthame_Schmeiser.04,Filbet_Perthame, Filbet} and for cells in a tissue \cite{Hillen.05,Chauviere_Hillen_Preziosi.07,loy2021EJAM, loy2019JMB,loy2020JMB}. In fact, the typical migration mode of a cell is the {\it run and tumble}, consisting in alternating runs over straight lines and reorientations, that may be biased by the presence of external signals affecting the choice of the direction, such as, for example, chemicals and the cell population density itself. At the population, or aggregate, level, the latter process may give rise to a tactic dynamics such as chemotaxis and adhesion, respectively. The run and tumble process may be modelled as a microscopic stochastic process named {\it velocity-jump process} \cite{Stroock}. It is a Markovian processes that prescribes a \textit{transition probability} $T$ of choosing a new velocity and a \textit{frequency of reorientation}  $\mu$.  In particular, the transition probability $T=T[\cS]$ may be influenced by the presence of an external signal $\cS$, that may embody the presence of a chemoattractant or of the cell population density. The kinetic equation that implements a velocity-jump process of intensity $\mu >0$, that is a piecewise deterministic Markov process in which we consider a transition probability $T[\cS]$, may be written as
\begin{equation}\label{eq:cinetique}
    \partial_t f(t,\x,v,\hv) +\vb \cdot \nabla_\x f(t,\x,v,\hv)= \mu \big(\rho(t,\x)T[\cS](v,\hv)-f(t,\x,v,\hv)\big),
\end{equation} 
where $f=f(t,\x,v,\hv)$ is a probability density function describing the distribution of the particle located at position $\x \in \Omega \in \mathbb{R}^d$, moving with speed $v\in [0,U]$ along direction $\hv\in \mathbb{S}^{d-1}$, for each time $t>0$. We also use the notation $\vb$ for the microscopic velocity of the cells that is given by the vector $\vb=v\hv$. As $(v,\hv) \in \mathcal{V}:=[0,U] \times \mathbb{S}^{d-1}$, then $\vb \in B(0,U)$, that is compact in $\mathbb{R}^d$ and symmetric.
The function $\rho(t,\x)$ denotes the number density of cells in position $\x$ at time~$t$:
\begin{equation*}\label{def:rho}
    \rho(t,\x)=\int_0^U \int_{\mathbb{S}^{d-1}} f(t,\x,v,\hv) \, dv d\hv, \qquad (t,\x) \in \mathbb{R}_+ \times \Omega.
\end{equation*}
Formally, \eqref{eq:cinetique} is a kinetic equation with linear relaxation operator of BGK type.

Another important issue in modeling cell migration is the nonvanishing size of the cell that gives rise to nonlocality in the physical space (see e.g. \cite{Armstrong_Painter_Sherratt.06,Chen2020PTRSB}  and references therein). Specifically at the kinetic level, a nonlocal gradient of the chemoattractant $\cS$ sensed with a sampling radius has been introduced in~\cite{Othmer_Hillen.02,hillen2007DCDSB}. In \cite{loy2019JMB,loy2020JMB, conte2022BMB, conte2023SIAP} the authors propose some models in which $T$ depends on a fixed external signal $\cS$ and also consider the case in which $\cS$ is the cell density~$\rho$, thus mimicking adhesion.  In \cite{loy2019JMB,loy2020JMB} the authors derive the so called macroscopic models for the aggregate quantities defined as averaged quantities (the statistical moments of~$f$), showing that keeping the nonlocality at the aggregate description implies a strong nonvanishing advection term. 
In such models, aggregation and concentrations have been observed, both in the case of linear models (i.e. $T$ depends on~$\cS$),  and nonlinear ones (i.e. $T$ depends on~$\rho$). In particular, in \cite{loy2020KRM} a stability analysis of a model with adhesion is performed and pattern formation is shown. 

Pattern formation may be seen in this context as a formation of small concentrations, typically persisting in time. As a matter of fact, this kind of solution may be represented as the sum of Dirac masses, in the form
\[
\rho_\varepsilon(t,\x) \approx \sum \bar{\rho}_i(t) \delta(x-\bar{\x}_i(t)),
\]
where $\varepsilon$ is a small parameter determining a specific regime, typically of high frequencies, in which there may be  formation of concentrations and patterns, $\bar{\x}_i$ is the location of a concentration point (that is a maximum point of the solution $\rho_\varepsilon$) and $\bar{\rho}_i$ is the weight of the concentration.
 A popular tool for analysing such concentration profiles is the real WKB ansatz or Hopf-Cole transform.
The leading idea is similar  to approximate the Dirac mass by Gaussians
\[
\delta(\x-\bar{\x}) \approx \dfrac{1}{\sqrt{2\pi \varepsilon}} \exp^{\dfrac{-|\x-\bar{\x}|^2}{2\varepsilon}}=\exp^{\dfrac{-|\x-\bar{\x}|^2-\varepsilon 2\pi \ln (\varepsilon)}{2\varepsilon}}.
\]
Therefore, the assumption is to consider a real WKB ansatz for the high frequency regime
(or Hopf-Cole transform) in the form
\[
\rho_\varepsilon(t,\x)=\exp^{-\dfrac{\varphi_{\varepsilon}(t,\x)}{\varepsilon}}, \qquad \lim_{\varepsilon\to 0} \varphi_\varepsilon \ge 0,
\]
where the concentration points are understood as the minima of the phase $\varphi_\varepsilon$. 
This kind of analysis typically leads, in the limit $\varepsilon \rightarrow 0$, to a constrained Hamilton-Jacobi equation for the phase. Such an Hamilton-Jacobi equation enables a rigorous derivation of the so-called canonical equation for the evolution of the maxima, which has
been formally proposed in the framework of adaptive dynamics in structured (by a trait) population equations to describe the trait evolution. In the context of adaptive dynamics, the maximum is interpreted as the `fittest trait', \cite{BaPe2015,AL_SM_BP,LLP2023,Lorenzi_2020} and the Hamilton-Jacobi equation also allows to find an equation for the location of the maxima, that is typically given by the roots of the equation obtained by setting the growth parameter to zero.

In the context of kinetic equations, this kind of ansatz has been first adopted by Bouin and Calvez in \cite{BC}, where they study a BGK model with a relaxation Maxwellian that is spatially homogeneous. In a high frequency regime, they assume a Hopf-Cole transform of the distribution   
\begin{equation}\label{def:HopfCole}
    f_\varepsilon(t,\x,v,\hv)= \exp^{-\dfrac{\varphi_\varepsilon(t,\x,v,\hv)}{\varepsilon}}
\end{equation}
and also assume a perturbed test function $\phi_\varepsilon(t,x,v,\hv)=\phi(t,x)+\varepsilon \eta(t,\x,v,\hv),$
i.e. the leading order in $\varepsilon$ only depends on $(t,\x)$.
The authors in~\cite{BC} derive in the limit $\varepsilon \rightarrow 0^+$ a Hamilton-Jacobi equation to which the potential is a viscosity solution. This has been extended in \cite{Caillerie} in order to take into account higher space dimensions, where the Hamiltonian may lack $\mathcal{C}^1$ regularity. In~\cite{bouin2015KRM,bouin2015ARMA,bouin2019EJAM} the authors study front propagation in transport-reaction kinetic equations. In~\cite{bouin2023JLMS} the authors study a BGK type equation in a high frequency regime and with a Maxwellian with vanishing variance and derive a new constrained nonlocal Hamilton-Jacobi equation.

In the present work, we want to analyse nonlocal kinetic equations of the same class of models as introduced in \cite{loy2019JMB, loy2020KRM} with the real WKB ansatz and the perturbed test function. The leading idea is to explain the concentration profiles by means of an appropriate Hamilton-Jacobi equation. The latter is derived from the kinetic equation after assuming a Hopf-Cole transform in the same spirit as in \cite{BC} and interpreting the location of the maxima (that is a spatial variable) as the fittest trait, like in adaptive dynamics.

In the first section, we will present the nonlocal kinetic equation under study along with the evolution equations for the aggregate quantities that can be derived in a regime of high frequencies and that can be obtained in a local and in a nonlocal rescaling. In section 3, we shall perform the WKB analysis of a linear nonlocal equation and present the concentration result as well as the canonical equation for the evolution of the maxima and some examples along with some numerical test. In section 4 we shall extend formally the analysis to a nonlinear case. In section 5 we draw some conclusion.

\section{Preliminaries}

\subsection{A nonlocal kinetic equation}

In the same spirit as~\cite{loy2019JMB},  we consider the kinetic equation~\eqref{eq:cinetique} with transition probability that depends on the external signal $\cS: \, \Omega \rightarrow \mathbb{R}_+$, that is measured nonlocally in the physical space and that affects the choice of the direction of the cells. The transition probability $T[\cS](v,\hv)$, that is a probability on $\mathcal{V}$ and depends on $\x$ through $\cS$, is in general defined by
\begin{equation}\label{def:T}
    T[\cS](v,\hv)=c(\x)\psi(v|\hv) b(\cS(\x+R\hv)),
\end{equation}
where $b(\cdot)$ is a function that weights the external field $\cS$, while the quantity~$R$ is the sensing radius defining the neighborhood of the particle where the field $\cS$ is measured.
The function $\psi=\psi(v|\hv): [0,U]\rightarrow \mathbb{R}_+$ is the probability density function of the possible speeds $v\in [0,U]$ on a given direction~$\hv$, satisfying
\[
\int_0^U\psi(v|\hv) \, dv =1 \quad \forall \hv \in \mathbb{S}^{d-1}.
\]
We denote its average speed (along direction $\hv$) $V_\psi$ and second statistical moment $D^2$ (that we assume to be independent of the direction), i.e. they are defined by
\begin{equation} \label{def:Vpsi}
    V_\psi(\hv):= \int_0^U\psi(v|\hv)v \, dv, \qquad D^2(\hv):= \int_0^U\psi(v|\hv)v^2 \, dv.
\end{equation}
The function $c(\x)$ is a normalization function defined by
\begin{equation*}\label{def:c}
    c(\x)^{-1} := \int_{\mathbb{S}^{d-1}} b(\cS(\x+R\hv)) \, d\hv.
\end{equation*}
This ensures that $T[\cS](v,\hv)$ is a probability density function on $\mathcal{V}=[0,U]\times \mathbb{S}^{d-1}$ as
\[
\int_0^U \int_{\mathbb{S}^{d-1}} T[\cS](v,\hv)  \, d\hv dv=1,
\]
in such a way that the number density is conserved at $(t,\x)$.
Then, we can also define the average velocity of the transition probability $T[\cS]$ as 
\begin{equation} \label{defU}
    \Ub_\cS(\x)=\int_0^U \int_{\mathbb{S}^{d-1}} T[\cS](v,\hv) \vb \, d\hv dv,
\end{equation}
and its variance-covariance matrix
\begin{equation}\label{defD}
\mathbb{D}_\cS(\x)=\int_0^U \int_{\mathbb{S}^{d-1}} T[\cS](v,\hv) (\vb-\Ub_{\cS})\otimes (\vb - \Ub_{\cS}) \, d\hv dv.
\end{equation}

\noindent We remark that, when $\Omega$ is bounded, in order to deal with the boundary, we must restrict the sensing radius using the formula
\begin{equation}\label{def:Rxv}
    R(\x,\hv):= \min \lbrace \lambda \in [0,R] : \x+\lambda\hv \in \Omega\rbrace .
\end{equation}
Eq.~\eqref{eq:cinetique} needs to be coupled with
initial and boundary conditions, defined by, respectively
\begin{equation}\label{eq:ci}
f(0,\x,v,\hv)=f^0(\x,v,\hv), \qquad  (\x,v,\hv)\in \Omega \times [0,U] \times \mathbb{S}^{d-1},
\end{equation}
\begin{equation}\label{eq:cb}
    f(t,\x,v,\hv)= \mathcal{R}[f_{|\Gamma_+}](t,\x,v,\hv) \qquad \x \in \partial \Omega, \, v\in [0,U], \, \hv \in \Gamma_-(\x),
\end{equation}
where
$$
\Gamma_{\pm}(\x):=\lbrace \hv \in \mathbb{S}^{d-1}: \hv\cdot \boldsymbol{n}(\x) \gtrless
 0 \rbrace,
$$
with ${\boldsymbol{n}}(\x)$ the outward normal to the boundary $\partial \Omega$ in the point $\x$. As boundary conditions for the kinetic equation, we assume a standard diffusive boundary condition \cite{Lods,Plaza} called Maxwellian boundary conditions, defined as
\begin{equation}\label{Max_cb}
\mathcal{R}[f_+](t,\x, v,\hat{v})=\alpha(\x)f(t,\x, v,\mathcal{W}(\hv))+(1-\alpha(\x))M(\x,v,\hv)\int_0^U\int_{ \hat{v}^*\cdot \boldsymbol{n}\ge 0} f(t,\x,v^*,\hat{v}^*)|\hv^*\cdot \boldsymbol{n}|d\hv^*  dv^*, 
\end{equation}
where $\mathcal{W}(\hv)=-\hv$ for the bounce back reflection condition and $\mathcal{W}(\hv)=\hv-2(\hv\cdot{\boldsymbol{n}}){\boldsymbol{n}}$ for the specular reflection.
Diffusive boundary conditions  are no-flux boundary conditions at the macroscopic level~\cite{Plaza}, in the sense that the total mass is conserved in $\Omega$. In fact, it may be proved that \cite{Plaza} if $f$ is a solution to~\eqref{eq:cinetique}-\eqref{eq:ci}-\eqref{eq:cb}-\eqref{Max_cb}, then the following is satisfied
\begin{equation}\label{noflux}
\int_0^U \int_{\mathbb{S}^{d-1}} f(t,\x,v,\hv)\vb\cdot {\bf n}(\x)d\hv \, dv=0, \quad \forall \x \in \partial \Omega, \quad t>0.
\end{equation}

The equilibrium distribution of~\eqref{eq:cinetique} is given by
\begin{equation*} \label{equilibrium}
    f^{\infty} (\x,v,\hv)=\rho(\x) T[\cS](v,\hv).
\end{equation*}
With classical arguments (Jensen's inequality) and assuming Maxwellian boundary conditions (that are nonabsorbing boundary conditions \cite{Cercignani}) it is easy to see that given a convex function $\Phi$, then
\[
\dfrac{d}{dt} \int_{\Omega}\int_0^U \int_{\mathbb{S}^{d-1}} \Phi\Big(f(t,\x,v,\hv)\Big) \, dv d\hv d\x \le  0,
\]
and the equality holds if and only if $f=f^\infty$. This equilibrium is asymptotically stable and does not depend on the initial condition.
As $T[\cS]$ depends on $\x$ through $\cS$, in order for $f^{\infty}$ to be a stationary equilibrium, then the following must be satisfied
\begin{equation}\label{eq:infty}
-\dfrac{\vb \cdot \nabla_\x \rho}{\rho}=\dfrac{\vb \cdot\nabla_\x T[\cS]}{T[\cS]}.  
\end{equation}


\subsection{Rescaling }\label{sec:aggregate}

We now consider a regime in which reorientations occur at random exponential times with rate $\mu=\dfrac{1}{\varepsilon}$, i.e., the dynamics is ruled by
\begin{equation}\label{eq:cinetique_hyp}
\partial_t f_\varepsilon(t,\x,v,\hv) +\vb \cdot \nabla_\x f_\varepsilon(t,\x,v,\hv)= \dfrac{1}{\varepsilon} \left(\rho_\varepsilon(t,\x)T[S]_\varepsilon(v,\hv)-f_\varepsilon(t,\x,v,\hv)\right) 
\end{equation}
where the limit $\varepsilon \rightarrow 0$ defines a high frequency regime.

On a one hand, a possible interpretation is to see Eq.~\eqref{eq:cinetique_hyp} as the result of a hyperbolic scaling of Eq.~\eqref{eq:cinetique} defined by
\begin{equation}\label{def:hyp}
(t,v,\x) \rightarrow  (\dfrac{t}{\varepsilon},v,\dfrac{\x}{\varepsilon}),   
\end{equation} 
that defines a long time scale (the equilibrium is reached fast) and a macroscopic (or large) space scale in which the interactions are localized. In fact, in this rescaling we also need to consider a scaling of the sensing radius $R$, i.e.
\begin{equation*}
    R \rightarrow \varepsilon R,
\end{equation*}
that naturally leads to a localization of the interactions of the cells with the background $\cS$. In this sense the large scale limit of~\eqref{eq:cinetique_hyp} for $\varepsilon\rightarrow 0$ leads to the hydrodynamic (fluid) behavior of the system on a macroscopic space scale that must be observed on a long time scale.

\noindent However, a priori we can consider the perspective of the following nondimensionalization
\begin{equation}\label{def:nondim}
    x \rightarrow \dfrac{x}{L}, \quad t \rightarrow \dfrac{t}{t_0}, \quad v \rightarrow \dfrac{v}{V}, \quad \rho \rightarrow \dfrac{\rho}{\bar{\rho}}, \quad f \rightarrow \dfrac{f}{\bar{\rho}/V^d}, \quad  T[\cS]\rightarrow \dfrac{T[\cS]}{V^d},
\end{equation}
where $t_0$ and $L$ are characteristic time and length scales of the system, $V$ is the typical speed, while $\bar{\rho}$ is a reference density.
Plugging~\eqref{def:nondim} in~\eqref{eq:cinetique} we obtain
\begin{equation}\label{eq:cinetique_nondim}
\textrm{St} \, \partial_t f(t,\x,v,\hv) +\vb \cdot \nabla_\x f(t,\x,v,\hv)= \dfrac{1}{\textrm{Kn}} \left(\rho(t,\x)T[S](v,\hv)-f(t,\x,v,\hv)\right),    
\end{equation}
where the kinetic Strouhal number $\textrm{St}$ and Knudsen number $\textrm{Kn}$ are defined as
\begin{equation*}
    \textrm{St}:=\dfrac{L}{V t_0}, \qquad \qquad 
    \textrm{Kn}:=\dfrac{V}{L \mu}.
\end{equation*}
The regime under consideration in~\eqref{eq:cinetique_hyp} corresponds to having parameters given by 
\begin{equation}\label{def:nondim_hyp}
     \textrm{St} \sim \mathcal{O}(1), \qquad \qquad  \textrm{Kn} \sim \mathcal{O}(\varepsilon) \ll 1 .
\end{equation}
By looking at~\eqref{eq:cinetique_nondim} we observe that the parameters regime~\eqref{def:nondim_hyp} corresponds to a large time horizon~$t_0$ satisfying
\begin{equation}\label{St_1_t0}
     t_0 \sim \dfrac{\varepsilon^{-1}}{\mu}.
\end{equation}
This may be rephrased saying that we choose $\varepsilon$ such that
\begin{equation}\label{def:eps}
    \dfrac{V}{L} \sim \mathcal{O}(\varepsilon)
\end{equation}
 and we choose a drift long time scale as
\begin{equation*}
    t_0=\dfrac{L}{V},
\end{equation*}
that satisfies~\eqref{St_1_t0} because of
\eqref{def:eps}. 
We remark that if $V=\mathcal{O}(1)$, then $\dfrac{1}{L}\sim \varepsilon$, that amounts to~\eqref{def:hyp} (where we use again $R\rightarrow \varepsilon R$). 
\\

On the other hand, if $L=\mathcal{O}(1)$, i.e., we observe the dynamics on the \textit{microscopic} space scale, then~\eqref{def:eps} amounts to a regime of very small speeds $V\sim \varepsilon$ that must be observed, in order to balance the smallness of the speed, on a long time scale. This can be seen as a scaling in the form
\begin{equation}\label{def:hyp_v}
(t,v,\x) \rightarrow  (\dfrac{t}{\varepsilon},\varepsilon v,\x).
\end{equation}
The latter may also be seen as a nonlocal regime as the sensing radius $R$ is not rescaled.

\subsection{Aggregate behaviour}

In the regime defined by~\eqref{eq:cinetique_hyp}, we may find limiting equations for the averaged population quantities. The single conservation law induces that the aggregate quantity is the mass and we obtain an evolution equations for the number density $\rho$.
Taking into account the equilibrium $T[\cS]_\varepsilon$, the formal expansion  of~\eqref{eq:cinetique_hyp} at order $\mathcal{O}(\varepsilon)$ is a diffusion-advection equation with a dominating drift term and small diffusivity
\begin{equation}\label{eq:macro}
 \partial_t \rho_\varepsilon +\nabla_\x \cdot (\rho_\varepsilon \Ub_{\cS}^\varepsilon)=\varepsilon \nabla_\x \cdot  \left(\nabla_\x \cdot( \mathbb{D}_\cS^\varepsilon \rho_\varepsilon)+\rho_\varepsilon \Ub_\cS^\varepsilon \nabla_\x \cdot \Ub_\cS^\varepsilon\right),
\end{equation}
where $\Ub_\cS^\varepsilon$ is the average of $T[\cS]_\varepsilon$ and $\mathbb{D}_\cS^\varepsilon$ its variance-covariance matrix as defined by~\eqref{defU} and~\eqref{defD}.   
The boundary conditions can be found by imposing~\eqref{noflux} to $f_\varepsilon$ \cite{Plaza} and this gains
\begin{equation}\label{eq:cb_macro}
    (\rho_\varepsilon \Ub_\cS^\varepsilon -\varepsilon\big(( \mathbb{D}_\cS^\varepsilon\nabla \rho_\varepsilon)+\rho_\varepsilon(\nabla \cdot\mathbb{D}_\cS^\varepsilon+ \Ub_\cS^\varepsilon \nabla \cdot \Ub_\cS^\varepsilon) \big)\cdot \boldsymbol{n}=0 \qquad \text{on } \; \partial \Omega.
\end{equation}

In the asymptotic limit $\varepsilon \rightarrow 0^+$, the dynamics is ruled by the equilibrium of~\eqref{eq:cinetique_hyp} at order zero in $\varepsilon$, that is defined by
\begin{equation*}\label{def:T0}
   T[\cS]_0:= \lim_{\varepsilon \rightarrow 0^+} T[\cS]_\varepsilon, 
\end{equation*}
in such a way that the evolution equation for $\rho=\rho_0$ is 
\begin{equation}\label{eq:macro_0}
 \partial_t \rho +\nabla_\x \cdot (\rho \Ub_{\cS}^0)=0.
\end{equation}
The boundary conditions, that can be derived substituting $f_\varepsilon$ in~\eqref{noflux} and letting $\varepsilon \rightarrow 0$, are given by 
\begin{equation}\label{eq:macro_cb}
    \rho \Ub_\cS^0(\x) \cdot \boldsymbol{n}(\x)  =0, \quad \text{for $\x\in \partial \Omega$},
\end{equation}
that are actually no-flux conditions for the conservation law~\eqref{eq:macro_0}.
As explained in Appendix~\ref{Appendix_cb}, the latter are actually only to be imposed on the entering region, but this can be in fact derived by the underlying kinetic boundary conditions that are imposed on the entering zone $\Gamma_-(\x)$. 
\\

In conclusion, both rescalings, the local macroscopic one~\eqref{def:hyp} or the nonlocal microscopic one~\eqref{def:hyp_v}, are possible and lead to the same equation~\eqref{eq:macro_0}, but the scales of the sensing radius differ and, as a consequence, $\Ub_\cS^0$ differs. For example, in the case $b(\cS)=\cS$, when the rescaling~\eqref{def:hyp} is performed (assuming $\cS$ smooth), then $T[\cS]_\varepsilon(v,\hv)=c(\x) \psi(v|\hv) \cS_\varepsilon(\x+\varepsilon R \hv)$, so that
$$
T[\cS]_0(v,\hv)=\psi(v|\hv)\dfrac{1}{|\mathbb{S}^{d-1}|}, \quad \Ub_\cS^0=\int_{\mathbb{S}^{d-1}} V_\psi(\hv) \, \hv d\hv, \quad \mathbb{D}_\cS^0=D^2\mathbb{I}.
$$
If $V_\psi(\hv)$ is even  (e.g., $V_\psi$ constant), then $\Ub_\cS^0=0$, so that from~\eqref{eq:macro} $\partial_t \rho_0=0$. Anyway, localization does not imply a vanishing drift in all cases. Let us, for example, consider a comparative sensing \cite{loy2019JMB}, which means that we assume that the turning rate depends on what is measured in $\x +R\hv$ and $\x -R\hv$, i.e.
\begin{equation*}\label{b.grad.rep}
b\big(\mathcal{S}(\x+\lambda\hv),\mathcal{S}(\x - \lambda\hv)\big) = \alpha+\beta  \dfrac{\mathcal{S}(\x-\lambda \hv) -\mathcal{S}(\x+\lambda \hv)}{2k+\mathcal{S}(\x+\lambda \hv) +\mathcal{S}(\x-\lambda \hv)}\,.
\end{equation*}
Then, assuming fast adaptation $\beta\rightarrow\frac \beta \varepsilon$, we find
\begin{equation*}\label{F.expansion}
T[\cS]_0(v,\hv)= \psi(v|\hv) \left(\alpha +\beta \dfrac{R\nabla \mathcal{S}(\x)\cdot \hv}{k+\mathcal{S}(\x)} \right) \, ,
\end{equation*}
which means that, even in the localized interactions regime, if $T[\cS]$ is given by a comparative sensing, the equilibrium depends on the directional derivative of the external field $\cS$ along each microscopic direction $\hv$. 
Conversely, in the regime~\eqref{def:hyp_v} we have, 
\[
T[\cS]_0=T[\cS], \quad \text{and thus,} \quad  \Ub_\cS^0=c(\x)\int_{\mathbb{S}^{d-1}} V_\psi(\hv) \cS(\x+R\hv) \, \hv d\hv,
\]
that is in general a nonvanishing quantity at the microscopic space scale.
In fact, we can remark that even in the case $V_\psi(\hv)=V_\psi$, then 
\[
\Ub_\cS^0=V_\psi c(\x)\int_{\mathbb{S}^{d-1}}  \cS(\x+R\hv) \, \hv d\hv
\]
is a nonvanishing quantity unless $\cS$ is spatially homogeneous. Therefore, the dominating drift term is due to the spatial heterogeneity that is sensed nonlocally. 
\\

In conclusion, Eq.~\eqref{eq:macro_0}, if derived as a large scale limit in the regime~\eqref{def:hyp}, has to be meant as a hydrodynamic limit on a macroscopic space scale in which interactions are localized and the (now local) equilibrium is reached fast as a longtime scale is observed. Conversely, it is derived in the regime~\eqref{def:hyp_v}, it implies a high frequency and small speeds regime on the microscopic space scale and slow time scale that is the same as the one of the original kinetic equation~\eqref{eq:cinetique}. Therefore, when derived in this regime, Eq.~\eqref{eq:macro_0} describes the evolution of the average number density on the original (microscopic) phase space.

\subsubsection{Diffusive limit}
When dealing with cell migration modeling, a typical rescaling is the diffusive one. In the present framework, it corresponds to choosing in the nondimensionalization a diffusive long time scale $t_0=\dfrac{L^2}{V^2}$ that actually satisfies $t_0=\dfrac{\varepsilon^{-2}}{\mu}$, in such a way that $\textrm{St}=\varepsilon$.
Therefore, the rescaled kinetic equation is in the form
\begin{equation}\label{eq:kin_resc_diff}
    \partial_t f_\varepsilon+\dfrac{1}{\varepsilon}\vb\cdot \nabla_\x f_\varepsilon= \dfrac{1}{\varepsilon^2}(\rho_\varepsilon T[\cS]_\varepsilon - f_\varepsilon).
\end{equation}  
When we consider a nonlocal diffusive rescaling, i.e.
\begin{equation}\label{def:diff_v}
(t,v,\x) \rightarrow  (\dfrac{t}{\varepsilon^2},\varepsilon v,\x)
\end{equation}
and $T[\cS]$ depends on $x$ and on $(v,\hv)$, then typically $\Ub_\cS$ is not a vanishing quantity and the aggregate equation for $\rho$ is
    \begin{equation}\label{eq:diff.2}
        \partial_t \rho+\nabla_\x \cdot (\Ub_\cS\rho)=\nabla \cdot \nabla\cdot\left(\rho\mathbb{D}_\cS\right).
    \end{equation}
    Conversely, when we consider a localized diffusive scaling, i.e.
\begin{equation}\label{def:diff}
(t,v,\x) \rightarrow  (\dfrac{t}{\varepsilon^2}, v,\dfrac{\x}{\varepsilon})    
\end{equation}
we can typically consider a Hilbert expansion for $\rho_\varepsilon$ and $T[\cS]_\varepsilon$, i.e.
\[
T[\cS]_\varepsilon=T[\cS]_0+\varepsilon T[\cS]_1, \qquad \int_{\mathcal{V}} T[\cS]_0 dv d\hv =1, \quad \int_{\mathcal{V}} T[\cS]_1 dv d\hv =0.
\]
When the solvability condition $\Ub_\cS^0=0$ is met, the macroscopic equation for $\rho=\rho_0$ is
\begin{equation}\label{eq:diff.1}
        \partial_t \rho+\nabla_\x \cdot (\Ub_\cS^1 \rho)=\nabla \cdot \nabla\cdot\left(\rho\mathbb{D}_\cS^0 \right).
\end{equation}
Then, supposing that
\begin{equation}\label{small_diff}
\mathbb{D}_\cS^0=\nu^2 \bar{\mathbb{D}}_\cS^0,
\end{equation}
when $\nu^2=\varepsilon$, i.e. for a small diffusivity, we essentially recover~\eqref{eq:macro} in the macroscopic limit.

\subsection{Limit for small $R$}
Let us define the characteristic length of variation of $\cS$ as
\begin{equation}\label{def:l_S}
l_\cS:= \dfrac{1}{\max \dfrac{|\nabla_\x \cS|}{\cS}}.
\end{equation}
We remark that when $R \ll l_\cS$, then we may consider the Taylor expansion of $\cS$ at first order
\begin{equation*}
    \cS(\x+R\hv)=\cS(\x)+R\hv \cdot \nabla_\x \cS(\x) + \mathcal{O}(R^2),
\end{equation*}
that is a positive quantity. Then, for example, in the case $b(\cS)=\cS$, we may approximate the probability density function $T[\cS]$ as
\begin{equation*}
    T[\cS](v,\hv)=\dfrac{\psi(v|\hv)}{|\mathbb{S}^{d-1}|}\left[1+R\hv \cdot \dfrac{\nabla_\x  \cS}{\cS}\right].
\end{equation*}
Then, choosing $L=l_\cS$ the nondimensionalization of~\eqref{eq:cinetique} leads to
\begin{equation*}
\textrm{St} \, \partial_t f_\varepsilon(t,\x,v,\hv) +\vb \cdot \nabla_\x f_\varepsilon(t,\x,v,\hv)= \dfrac{1}{\textrm{Kn}} \left(\rho_\varepsilon(t,\x)\dfrac{\psi(v|\hv)}{|\mathbb{S}^{d-1}|}(1+\eta \hv \cdot \dfrac{\nabla_\x  \cS(\x)}{\cS(\x)} )-f_\varepsilon(t,\x,v,\hv)\right),    \end{equation*}
where
\begin{equation*}
    \eta:=\dfrac{R}{l_\cS}.
\end{equation*}
 With the choice $\dfrac{1}{L}\sim \varepsilon$, that amounts to~\eqref{def:hyp}, then we obtain $\eta=R \varepsilon$ and
\begin{equation*}
\partial_t f_\varepsilon(t,\x,v,\hv) +\vb \cdot \nabla_\x f_\varepsilon(t,\x,v,\hv)= \dfrac{1}{\varepsilon} \left(\rho_\varepsilon(t,\x)(T[S]_0+\varepsilon  T[\cS]_1)-f_\varepsilon(t,\x,v,\hv)\right), 
\end{equation*}
where $T[\cS]_0(v,\hv)=\dfrac{\psi(v|\hv)}{|\mathbb{S}^{d-1}|}$ and $T[\cS]_1(v,\hv)=\dfrac{\psi(v|\hv)}{|\mathbb{S}^{d-1}|}R \hv \cdot \dfrac{\nabla_\x  \cS(\x)}{\cS(\x)}$. This corresponds to considering~\eqref{eq:cinetique} with~\eqref{def:hyp}, as the leading order term is local while the role of the sensing radius enters the dynamics as a higher order term.
In this case the evolution equations for $\rho_\varepsilon$ correspond in the macroscopic point of view~\eqref{def:hyp} and in the high frequency (microscopic) one~\eqref{def:hyp_v}, i.e.,
\begin{equation}\label{eq:diffRsmall}
    \partial_t \rho_\varepsilon = \varepsilon \Delta \rho_\varepsilon.  
\end{equation}
When choosing the diffusive scaling (choosing $V_\psi$ constant for simplicity) we obtain~\eqref{eq:diff.1} with $\mathbb{D}_\cS^0=\mathbb{I}$ and $\Ub_\cS^1=R\dfrac{\nabla_x \cS}{\cS}$, that is the Keller and Segel model~\cite{loy2019JMB,Keller_Segel}.


\section{Concentration profile and the Hamilton–Jacobi equation}

We want to study the concentration profile of the solution $f_\varepsilon$ of Eq.~\eqref{eq:cinetique_hyp} by studying the equation for a potential $\varphi_\varepsilon$ obtained through the Hopf-Cole transform~\eqref{def:HopfCole}. 
We expect that $\varphi_\varepsilon(t,\x,v,\hv)$ behaves like a quadratic and thus that $f_\varepsilon(t,\x,v,\hv)$ behaves like a Dirac mass near each concentration point. For that reason, we study the limit of $\varphi_\varepsilon$.

\subsection{The Hamilton–Jacobi equation}

At first we remark that, from Eq.~\eqref{eq:cinetique_hyp},  $\varphi_\varepsilon$ satisfies the equation
\begin{equation}\label{eq:phia}
    \partial_t \varphi_\varepsilon+\vb \cdot \nabla_{\x} \varphi_\varepsilon=1-T[S]_\varepsilon(v,\hv)\int_{\mathcal{V}} \exp^{\dfrac{-\varphi_\varepsilon(t,\x,w,\hw)+\varphi_\varepsilon(t,\x,v,\hv)}{\varepsilon}} \, dw d\hw .
\end{equation}
From this we get
\begin{equation} \label{eq:phi}
    \left(1-\partial_t \varphi_\varepsilon-\vb \cdot \nabla_{\x} \varphi_\varepsilon\right)= T[S]_\varepsilon(v,\hv)\int_{\mathcal{V}} \exp^{\dfrac{-\varphi_\varepsilon(t,\x,w,\hw)+\varphi_\varepsilon(t,\x,v,\hv)}{\varepsilon}} \, dw d\hw.
\end{equation}
Following \cite{BC, LLP2023}, we may also look for $f_\varepsilon$ under the form
\begin{equation}  \label{defQphieps}
    f_\varepsilon (t,\x,v,\hv) =Q_\varepsilon(t,\x,v,\hv) \exp^{-\dfrac{\tilde \varphi_\varepsilon(t,\x)}{\varepsilon}}, \qquad \varphi_\varepsilon(t,\x,v,\hv)= \tilde \varphi_\varepsilon(t,\x) -\varepsilon \ln Q_\varepsilon(t,\x,v,\hv),
\end{equation}
with $\tilde \varphi_\varepsilon$ and $Q_\varepsilon$ to be determined. Setting 
\begin{equation}  \label{defpHieps}
p_\varepsilon=\nabla_\x \tilde \varphi_\varepsilon, \qquad H_\varepsilon=-\partial_t \tilde \varphi_\varepsilon,
\end{equation}
we can write Eq.~\eqref{eq:phi} as
\begin{equation}\label{eq:eigen_pbbis}
\begin{array}{ll}
\varepsilon [\partial_t Q_\varepsilon(t,\x,v,\hv)+\vb \cdot \nabla_{\x}Q_\varepsilon(t,\x,v,\hv)] +\! &(1 + H_\varepsilon - v\hv \cdot p_\varepsilon) Q_\varepsilon(t,\x,v,\hv)
\\[5pt]
&= \displaystyle T[\cS]_\varepsilon(v,\hv) \int_{\mathcal{V}} Q_\varepsilon(t,\x,w,\hw) d\hw dw . 
\end{array}
\end{equation}
The formal limit as $\varepsilon\to 0$ gives us 
\[
(1 + H - v\hv \cdot p) Q(t,\x,v,\hv)= \displaystyle T[\cS]_0(v,\hv) \int_{\mathcal{V}} Q(t,\x,w,\hw) d\hw dw .
\]
This can be interpreted as the eigenvalue-eigenfunction problem in $(v,\hv)$, with $(t,\x)$ parameters, which is to find $(H,\mathcal{Q}) $ such that
\begin{equation}\label{eq:eigen_pb}
(1 + H - v\hv \cdot p) \mathcal{Q}(\x,p, v,\hv) = T[\cS]_0(v,\hv) \int_{\mathcal{V}} \mathcal{Q}(\x,p, w,\hw) d\hw dw .
\end{equation}
Thanks to the Krein-Rutman theory, see \cite{LamLou},  with good properties of $T[\cS]_0$ to be discussed later, this eigenproblem has a unique solution once normalized as 
\begin{equation}\label{eq:eigen_pb_norm}
\mathcal{Q}(\x,p,v,\hv) >0, \qquad \int_{\mathcal{V}} \mathcal{Q}(\x,p,v,\hv) d\hv dv = 1, \qquad \forall p, \; \x. 
\end{equation}
The eigenvalue $H$ is solely determined by the parameters $p$ and $\cS(\x)$ and we can write $H=H(\x,p)$ which provides us with the Hamilton-Jacobi equation for the dominant term in~\eqref{defpHieps} 
\begin{equation}\label{eq:HJ}
\partial_t  \varphi + H(\x, \nabla_\x \varphi)=0,
\end{equation}
with $\varphi$ the (formal) common limit of $\varphi_\varepsilon$ or $\tilde \varphi_\varepsilon$
\begin{equation*}
   \tilde \varphi:=\lim_{\varepsilon \rightarrow 0^+} \tilde \varphi_\varepsilon .
\end{equation*}
Then, we recover the limiting corrector $Q(t,\x, v ,\hv)=\mathcal{Q}\big(\x,\nabla_\x \varphi (t,\x), v,\hv\big)$.
\\

Furthermore, adding the condition (assumed to hold initially)
\[
\int_{\mathcal{V}} Q_{\varepsilon}(t,\x,v,\hv) d\hv dv = 1, \qquad \forall t\geq 0, \; \x\in \Omega , 
\]
which in turn implies 
\[
\int_\Omega \exp^{-\dfrac{\tilde \varphi_\varepsilon(t,\x)}{\varepsilon}} d\x =1, \qquad \forall t,
\]
we can expect that the problem~\eqref{eq:eigen_pbbis} itself has a particular solution $Q_\varepsilon(t,\x,v,\hv)$ similar to the {\em principal bundle}, see \cite{LamLou, LLP2023}, for parabolic equations. It is similar to a time dependent eigenvalue problem.  Up to our knowledge this notion has never been studied for kinetic equations. We can expect it defines a time-dependent functional Hamiltonian $H_{\varepsilon} (t, \x, [\nabla_x \tilde \varphi_{\varepsilon}]) \to H(\x,\nabla_\x  \varphi (t,\x))$. This allows us to search for the solution of a functional Hamilton-Jacobi equation 
\begin{equation}\label{eq:HJfunc}
\partial_t \tilde \varphi_{\varepsilon}(t,\x)+ H_{\varepsilon}\big(t,\x, [\nabla_\x \tilde \varphi_{\varepsilon}]\big)=0.
\end{equation}
Then $f_\varepsilon (t,\x,v,\hv)$ in~\eqref{defQphieps} is an exact particular solution of Eq.~\eqref{eq:cinetique_hyp} solely determined by the initial concentration profile $\varphi_{\varepsilon}(0,\x)$. As $\varepsilon$ vanishes, this particular solution attracts all solutions with the same initial concentration profile $\varphi(0,\x)$.
\\

These formal conclusions rely on the possibility to define a smooth Hamiltonian $H(\x,p)$, a question we analyse now.

\subsection{The effective Hamiltonian}

As in \cite{BC}, one can characterize the eigenvalue $H(\x,p)$ arising in Eq.~\eqref{eq:eigen_pb} which can be written as
\begin{equation}\label{def:Q}
\mathcal{Q}(\x,p,v,\hv) = \dfrac{ T[\cS]_0(v,\hv) \int_{\mathcal{V}} \mathcal{Q}(\x,p, w,\hw) d\hw dw}{1 + H(\x, p) - v\hv \cdot p} >0.
\end{equation}
We remind that $H(\x,p)$ also depends explicitly on $\x$ as $T[\cS]_0$ depends on $\x$. 
Integrating with respect to $v,\hv$ and using~\eqref{eq:eigen_pb_norm},
we obtain the following problem: find $H$ such that 
\begin{equation}\label{eq:H}
    1=\int_{\mathcal{V}}  \dfrac{T[S]_0(v,\hv)}{1+H(\x, p)-v\hv \cdot p} \, dv d\hv.
\end{equation}
In particular we obviously have
\begin{equation*} 
    1= \int_{\mathcal{V}} \dfrac{T[S]_0(v,\hv)}{1+H(\x, 0)} \, dv d\hv, \qquad H(\x, 0)=0.
\end{equation*}
Eq.~\eqref{eq:H} can be uniquely solved by strict decay in $H$ and also gives that 
\[
-U |p| < H(\x, p) < U |p|,
\]
because when $H=U |p|$ the denominator is larger than $1$ for all $(v,\hv)$ and  when $H=- U |p|$ the denominator is smaller than $1$.

However these bounds are not enough to compute from this Hamiltonian a positive eigenfunction $\mathcal{Q}$. As observed in~\cite{Caillerie}, it is necessary to introduce some further assumption. We define the values of $H$ when the denominator vanishes as
\[
\underline H (p)= -1 + \max_{\mathcal{V}} \, [ v \hv\cdot p ].
\]
We then need to assume
\begin{equation} \label{as:H}
\inf_{\x} \int_{\mathcal{V}}  \dfrac{T[S]_0(v,\hv)}{1+\underline H (p) -v\hv \cdot p} \, dv d\hv > 1.
\end{equation} 
The latter ensures that $H(\x,p)>\underline H (p) $ and thus that the denominator, and therefore $\mathcal{Q}$, are positive for the solution of Eq.~\eqref{eq:H}. 
This integral blows-up in  1D and the condition is always satisfied. In higher dimension this restriction is needed. This is the so-called `dimensionality problem' as mentioned in~\cite{Caillerie}.

Additionally, differentiating~\eqref{eq:H} in $p$, we find (ignoring the dependence on $\x$ for simplicity) 
\[
0=\int_{\mathcal{V}}  \dfrac{T[S]_0(v,\hv)(\nabla_p H(p)-v\hv)}{(1+H(p)-v\hv \cdot p)^2} \, dv d\hv.
\]
Then, as we are interested in the minima points of $\varphi$ (that are the maxima of $\rho$), we look for the values of the Hamiltonian $H$ in $p=\nabla \varphi=0$, i.e., using the definition~\eqref{defU}, we get
\begin{equation} \label{Hp0}
    \nabla_p H (\x,0)=\Ub_{\cS}^0 (\x),
\end{equation}
which, in general, does not vanish, as already argued, as $T[\cS]_0$ is not in principle symmetric as a function of $\hv$, as instead assumed in \cite{BC}.
Moreover, differentiating twice, we find 
\[
\int_{\mathcal{V}}  \dfrac{T[S]_0(v,\hv)D^2H(p)}{(1+H(p)-v\hv \cdot p)^2} \, dv d\hv=2\int_{\mathcal{V}} \dfrac{T[S]_0(v,\hv)(\nabla_pH(p)-v\hv)\otimes (\nabla_pH(p)-v\hv) }{(1+H(p)-v\hv \cdot p)^3} \, dv d\hv,
\]
and then we have that $D^2_p H$ is positive definite as~\eqref{as:H} holds. We may also 
compute 
\begin{equation} \label{D2H0}
D^2_p H(x,0)= 2 \mathbb{D}_\cS(\x).    
\end{equation}

\subsection{The concentration result}

To simplify, we work in the full space, $\mathbb{R}^d\times \mathcal{V}$ instead of $\Omega\times \mathcal{V}$. We assume initially that uniformly in $\varepsilon$
\begin{equation} \label{as:init1}
\begin{cases}
 \varphi_\varepsilon^0 \in L_\infty, \qquad |\nabla_\x \varphi_\varepsilon^0| \in L_\infty, \qquad \varphi_\varepsilon^0(\x,v,\hv) = \tilde \varphi^0(\x)+O(\varepsilon), 
 \\
    \partial_t \varphi_\varepsilon^0 := -\vb \cdot \nabla_{\x} \varphi_\varepsilon^0+ 1-T[S]_\varepsilon(v,\hv) {\displaystyle \int_{\mathcal{V}}} \exp^{\dfrac{-\varphi_\varepsilon^0(\x,w,\hw)+\varphi_\varepsilon^0(\x,v,\hv)}{\varepsilon}} \, dw d\hw \in L_\infty,
\end{cases}\end{equation}
\begin{equation} \label{as:init2}
\int_{\mathbb{R}^d \times\mathcal{V}} f^0_\varepsilon (\x,v,\hv) d\x dv d\hv =1,  \qquad \int_{\mathbb{R}^d \times\mathcal{V}} |\x| f^0_\varepsilon (\x,v,\hv) d\x dv d\hv  \quad \text{is bounded},
\end{equation}
and for some constants $T_- >0$, $L_M>0$
\begin{equation} \label{as:TS}
\begin{cases}
 T[\cS]_\varepsilon \geq T_->0, \qquad T[\cS]_\varepsilon + | \nabla_x T[\cS]_\varepsilon |+ | \nabla_\vb T[\cS]_\varepsilon | \leq L_M,
\\[5pt]
T[\cS]_\varepsilon \rightarrow T[\cS]_0  \quad \text{uniformly}.
\end{cases}\end{equation}
Then we can prove the
\begin{theorem}
We make the assumptions\eqref{as:H} and~\eqref{as:init1}--\eqref{as:TS}. Then, after extractions,
\\
(i) $\varphi_\varepsilon $ is uniformly (in $\varepsilon$) bounded and Lipschitz (locally in time),
\\
(ii) $\varphi_\varepsilon $ converges locally uniformly on $\mathbb{R}_+\times \mathbb{R}^d \times \mathcal{V}$ toward $\varphi$ where $\varphi$ does not depend on~$v,\hv$. Moreover, $\varphi$ is the viscosity solution of the Hamilton-Jacobi Eq.~\eqref{eq:HJ} with initial condition $\varphi^0 (x)$ and with a convex Hamiltonian $H(\x, p)$ uniquely implicitly determined by the formula~\eqref{eq:H},
\\
(iii)  $f_\varepsilon $ converges weakly to a measure $f$ supported by $\{\varphi =0 \}$.
\end{theorem}

\noindent{\bf Remarks.} 1. Compared to \cite{BC}, the kernel $T[\cS]_\varepsilon$ depends on $\x$, which is an additional major technical difficulty. Also a difference here is the $\varepsilon$ dependency which is not relevant with our assumptions.
\\
2. The author in \cite{Caillerie} faces the difficulty of gradient estimates as here. He argues by limsup-liminf arguments which optimizes the assumptions. Here we do not go to this elaborate method and use simpler arguments based on Lipschitz estimates. 
\\
3. When $\Omega$ is bounded then we impose no-flux  boundary conditions \cite{Plaza} 
\[
\int_{\mathcal{V}} f_\varepsilon v\hv \cdot \boldsymbol{n} \, dv d\hv =0, \qquad \x \in \partial \Omega .
\]
Therefore, considering~\eqref{defQphieps}, we have
\[
\exp^{-\dfrac{\tilde{\varphi}_\varepsilon}{\varepsilon}}\int_{\mathcal{V}}Q_\varepsilon(t,\x,v,\hv) v\hv \cdot \boldsymbol{n} \, dv d\hv =0,\qquad \x \in \partial \Omega .
\]
and thus 
\[
\int_{\mathcal{V}} Q_\varepsilon(t,\x,v,\hv) v\hv \cdot \boldsymbol{n} \, dv d\hv=0, \qquad \x \in \partial \Omega ,
\]
and then in the limit $\varepsilon \rightarrow 0^+$
\begin{equation}\label{cb:HJ}
\int_{\mathcal{V}} \mathcal{Q}(x,p,v,\hv) v\hv \cdot \boldsymbol{n} \, dv d\hv=0, \qquad \x \in \partial \Omega .
\end{equation}
It would be interesting to investigate if this relation can be interpreted as a Neumann boundary condition on $p=\nabla_\x \varphi$.
Remark that for small $p$, Taylor expanding $\mathcal{Q}$ defined in~\eqref{def:Q} as a function of $p$ and plugging the expression into~\eqref{cb:HJ}, we find 
\begin{equation}\label{cb:HJ_smallp}
[\varphi \Ub_\cS^0+\mathbb{D}_\cS^0\nabla \varphi] \cdot \boldsymbol{n} =0.
\end{equation}
\\
4. The corrector satisfies, according to the Hopf-Cole transform \eqref{defQphieps}, approximately $\mathcal{Q}_\varepsilon =\dfrac{f_\varepsilon}{\rho_\varepsilon}$. However, we also have that $\dfrac{f_\varepsilon}{\rho_\varepsilon}\rightarrow T[\cS]_0$ and $\mathcal{Q}_\varepsilon \rightarrow \mathcal{Q}$ when $\varepsilon$ tends to zero. Actually, thanks to~\eqref{eq:eigen_pb}, we find $\mathcal{Q}(\x,0,v,\hv) = T[\cS]_0(v,\hv)$.


\bigskip

\begin{proof} The proof uses standard arguments, see for instance  \cite{crandall1992,GB:94}, and we only sketch it. We begin with standard a priori estimates (i) for the solution of Eq.~\eqref{eq:phia}. From assumption~\eqref{as:init1}, we infer
\[
|\partial_t \varphi_\varepsilon (t,x,v,\hv)| \leq |\partial_t \varphi_\varepsilon^0(x,v,\hv)| \leq C,
\]
therefore 
\[
| \varphi_\varepsilon (t,x,v,\hv)| \leq | \varphi_\varepsilon^0(x,v,\hv)| + Ct.
\]
Also, still using the maximum principle for derivatives and the already proved bounds (here Eq.~\eqref{eq:phia} is used again for $\x$-derivative of $T[S]$), we have that
\[
\sum_{i=1}^d  | \partial_{i} \varphi_\varepsilon (t,x,v,\hv)| \leq \sum_{i=1}^d | \partial_{i} \varphi^0_\varepsilon (x,v,\hv)| e^{Ct} +C t.
\]
Since this estimate is more elaborate, we prove it. Differentiating Eq.~\eqref{eq:phia} in $\x_i$, and setting $\psi_i= \partial_{i} \varphi_\varepsilon (t,x,v,\hv)$, we find
\[
\partial_t \psi_i+ \vb \cdot \nabla \psi_i  = -T[S]_\varepsilon\int_{\mathcal{V}} \exp^{\dfrac{-\varphi_\varepsilon(t,\x,w,\hw)+\varphi_\varepsilon(t,\x,v,\hv)}{\varepsilon}}\frac{\psi_i(v,\hv)- \psi_i(w,\hw)}{\varepsilon} \, dw d\hw + RHS
\]
where the RHS term is 
\[
RHS= \partial_{i}T[S]_\varepsilon(v,\hv) \int_{\mathcal{V}} \exp^{\dfrac{-\varphi_\varepsilon(t,\x,w,\hw)+\varphi_\varepsilon(t,\x,v,\hv)}{\varepsilon}}\, dw d\hw.
\]
Using again Eq.~\eqref{eq:phia}, it can be estimated as
\[\begin{array}{ll}
|RHS| &\leq \frac{|\partial_{i}T[S]_\varepsilon(v,\hv)|}{T[S]_\varepsilon(v,\hv)} T[S]_\varepsilon(v,\hv)  \int_{\mathcal{V}} \exp^{\dfrac{-\varphi_\varepsilon(t,\x,w,\hw)+\varphi_\varepsilon(t,\x,v,\hv)}{\varepsilon}}\, dw d\hw
\\[5pt]
&=\frac{|\partial_{i}T[S]_\varepsilon(v,\hv)|}{T[S]_\varepsilon(v,\hv)} [ 1-\partial_{t} \varphi_\varepsilon  -\vb\cdot\nabla \varphi_\varepsilon ],
\end{array} \]
and, thus, using the time derivative estimate, we conclude that
\[
|RHS| \leq Ct+ C \sum_{i=1}^d  |\psi_i|. 
\]
With this observation, we can use the maximum principle for $\psi_i e^{-Ct} +C$ and conclude the bounds on the $\x$ derivatives.

With these estimates, we conclude that for $t\leq T$, we have
\[\int_{\mathcal{V}} \exp^{\dfrac{-\varphi_\varepsilon(t,\x,w,\hw)+\varphi_\varepsilon(t,\x,v,\hv)}{\varepsilon}} \, dw d\hw \leq C(T),
\]
which tells us that a limit of $\varphi_\varepsilon$ depends only on $(t,\x)$. As in \cite{BC}, it also gives directly the last estimate of (i), that is
\[
| \partial_{\vb} \varphi_\varepsilon (t,x,v,\hv)| \leq \big[ | \partial_{\x} \varphi^0_\varepsilon (x,v,\hv)|+| \partial_{\vb} \varphi^0_\varepsilon (x,v,\hv)| \big] e^{Ct} +C t.
\]
\\

We are now in the same situation as \cite{BC} and the rest of the argument follows in a similar way. Using the perturbed test function method, \cite{evans1989}, we obtain the statement (ii) thanks to the assumption~\eqref{as:H} which allows us to handle $\mathcal Q.$
\\

Finally, for the statement (iii), we notice that the mass conservation is immediate. Then,  we observe that
\[
\frac{d}{dt} \int_{\mathbb{R}^d \times\mathcal{V}} |\x| f_\varepsilon (t,\x,v,\hv) d\x dv d\hv \leq  \int_{\mathbb{R}^d \times\mathcal{V}} |\vb| f_\varepsilon (t,\x,v,\hv) d\x dv d\hv \leq U.
\]
Therefore $f_\varepsilon$ is a tight probability measure and, after extraction, it converges weakly to a probability measure and the only possible concentration points are when $\varphi(t,\x)$ is zero (see \cite{BaPe2015, AL_SM_BP} for details and consequences).
\end{proof}

\subsection{An eikonal equation}
Given~\eqref{eq:macro}, in the limit $\varepsilon \rightarrow 0^+$, the phase $\phi_\varepsilon=-\varepsilon \log \rho_\varepsilon$ satisfies the Hamilton-Jacobi equation
\begin{equation}\label{eq:eikonal}
    \partial_t \phi+ \Ub_\cS^0 \cdot \nabla \phi+\textrm{tr}\left[ \mathbb{D}_\cS^0(\nabla_\x\phi\otimes \nabla_\x \phi)\right]=0.
\end{equation}
When working in a bounded domain, from~\eqref{eq:cb_macro}, we additionally obtain the boundary condition
\[
[\phi \Ub_\cS^0+\mathbb{D}_\cS^0\nabla \phi] \cdot \boldsymbol{n} =0,
\]
that is the same as~\eqref{cb:HJ_smallp}.
As $\varphi$ satisfies~\eqref{eq:HJ} and $\phi$ satisfies~\eqref{eq:eikonal}, we should be lead to conclude, as observed in \cite{BC}, that the two procedures (aggregate quantities limit and WKB analysis) do not commute in general, in particular because the Hilbert expansion is additive, while the Hopf-Cole one is multiplicative. However, Eq.~\eqref{eq:eikonal} may be seen as~\eqref{eq:HJ} where the quadratic expansion of the Hamiltonian $H$ in a neighborhood of $\nabla \phi=0$, that characterizes the minima points, is considered, remembering~\eqref{Hp0}-\eqref{D2H0}.
By exploiting this observation, it is possible to detect a regime in which the two procedures may commute.

Let us consider the regime~\eqref{eq:diff.2} or~\eqref{eq:diff.1} and the assumption of small diffusivity~\eqref{small_diff}.
Then, considering $\rho_\nu=\exp^{-\dfrac{\phi}{\nu}}$ and
letting $\nu \rightarrow 0$, 
Eq.~\eqref{eq:diff.2} becomes 
\begin{equation}\label{eq:eikonal_diff2}
\partial_t \phi+ \Ub_\cS \cdot \nabla \phi+\textrm{tr}\left[ \bar{\mathbb{D}}_\cS(\nabla_\x\phi\otimes \nabla_\x \phi)\right]=0,
\end{equation}
while Eq.~\eqref{eq:diff.1} becomes
\begin{equation}\label{eq:eikonal_diff1}
\partial_t \phi+ +\Ub_\cS^1\cdot \nabla \phi +\textrm{tr}\left[ \bar{\mathbb{D}}_\cS^0(\nabla_\x\phi\otimes \nabla_\x \phi)\right]=0.
\end{equation}

We now consider
\[
\varepsilon \rightarrow \varepsilon \nu
\]
and plug \eqref{def:HopfCole} in \eqref{eq:kin_resc_diff};
in the limit $\varepsilon \rightarrow 0$, we obtain
\begin{equation}\label{eq:HJ_nu}
\partial_t (\left[\dfrac{\varphi}{\nu}\right])+H_\nu (\nabla \left[\dfrac{\varphi}{\nu}\right])=0,
\end{equation}
where $H_\nu$ is implicitly defined by
\[
1=\nu\int_{\mathcal{V}}  \dfrac{T[S]_0(v,\hv)}{\nu+H(\x, p)-v\hv \cdot p} \, dv d\hv, \quad p= \nabla \left[\dfrac{\varphi}{\nu}\right].
\]
Then, as $ D^2 H(0)= \dfrac{2}{\nu^2} \mathbb{D}_\cS^0$, considering $\nu$ small and assuming the small diffusivity~\eqref{small_diff}, we obtain
\[
H(\x,\nabla \varphi)=H(\x,0)+  \nabla \varphi \cdot \nabla_p H(\x,0) + \dfrac{1}{2}\textrm{tr}\left[\nabla \varphi \otimes \nabla \varphi D_p^2 H(\x,0)\right]= \Ub_\cS^0 \cdot \nabla \varphi  +  \textrm{tr}\left[\bar{\mathbb{D}}_\cS^0\nabla \varphi \otimes \nabla \varphi \right].
\]
Plugging the latter in~\eqref{eq:HJ_nu} allows to obtain the equivalent form of Eq.~\eqref{eq:eikonal}.

Let us now consider a spatially homogeneous $T$ such that
\[
\int_\mathcal{V} T dv d\hv =1, \qquad \int_\mathcal{V} T v dv d\hv =0, \qquad \int_\mathcal{V} T v^2\hv\otimes \hv \, dv d\hv ^2 =\nu^2 \mathbb{I}.
\]
It is the case for example, in one-dimension, where $\mathcal{V}=[-U,U]$, and we choose
$
T=\exp^{-\dfrac{v^2}{2}}$, $ T=\dfrac{1}{U}$, or $ T=\dfrac{1}{2}\left(\delta(v-U)+\delta(v+U)\right)$, see \cite{BC}.
Then, performing the WKB analysis leads to ~\eqref{eq:HJ_nu} and, then, to $\partial_t \varphi +|\nabla^2 \varphi|=0$, while, starting from~\eqref{eq:diffRsmall} we obtain $\partial_t \phi +|\nabla^2 \phi|=0$, that is~\eqref{eq:eikonal_diff1} with $\Ub_\cS^1=0$. We highlight, in fact, that in these cases there is no correction term $T[\cS]_1$ (and, then no $\Ub_\cS^1$). In conclusion, the two procedures commute in the regime of small $\nu$.

Interestingly, in the case 
\[
T[\cS]=c(\x)\cS(\x+R\hv),
\]
considering \eqref{def:hyp_v} or \eqref{def:diff_v} and a large $R$, the two procedures lead to \eqref{eq:diff.2} and \eqref{eq:eikonal_diff2}, respectively. Conversely, if we consider a small $R$ and $T[\cS]_0, T[\cS]_1$ as defined as a consequence of a localized scaling \eqref{def:diff}, we obtain \eqref{eq:diff.1} with $\mathbb{D}_\cS^0=\mathbb{I}$, $\Ub_\cS^1=R\dfrac{\nabla \cS}{\cS}$. If in the WKB analysis we consider \eqref{def:hyp}, we obtain  the eikonal equation $\partial_t \varphi + |\nabla^2 \varphi|=0$ and the higher order effect is naturally lost in the localized hyperbolic scaling.
\commentout{\noindent \textbf{Remark.}
Given~\eqref{eq:macro}, in the limit $\varepsilon \rightarrow 0^+$, the phase $\phi_\varepsilon=-\varepsilon \log \rho_\varepsilon$ satisfies the Hamilton-Jacobi equation
\begin{equation}\label{eq:eikonal}
    \partial_t \phi+ \Ub_\cS^0 \cdot \nabla \phi+\textrm{tr}\left[ \mathbb{D}_\cS^0(\nabla_\x\phi\otimes \nabla_\x \phi)\right]=0.
\end{equation}
When working in a bounded domain, from~\eqref{eq:cb_macro}, we additionally obtain the boundary condition
\[
[\Ub_\cS^0+\mathbb{D}_\cS^0\nabla \phi] \cdot \boldsymbol{n} =0.
\]
As $\varphi$ satisfies~\eqref{eq:HJ} and $\phi$ satisfies~\eqref{eq:eikonal}, we conclude, as observed in \cite{BC}, that the two procedures (hydrodynamic limit and large deviations approach) do not commute. In fact the hydrodynamic limit generates a Hamiltonian which is the quadratic expansion of the Hamiltonian $H$ in~\eqref{D2H0} in a neighborhood of $\nabla \phi=0$, that characterizes the maxima points.}
\subsection{Dynamics of the concentration points}

Not only the Hamilton-Jacobi and eikonal equations give the microscopic shape of the Dirac concentration of solutions, when they occur, but it also allows to recover their dynamics. For this we look for the trajectory $\bar{\x}(t)$ of the maxima of~$\rho$. In the context of adaptive dynamics, this is interpreted as the `fittest trait', \cite{BaPe2015,AL_SM_BP,LLP2023}. First of all, we remark that in the regime~\eqref{eq:cinetique_hyp} for $\varepsilon \rightarrow 0$, each trajectory in the physical space (see~\eqref{eq:macro_0}) follows the differential equation
\begin{equation}\label{eq:traj}
    \dot{\x}(t)=\Ub_\cS^0(\x(t)).
\end{equation}
For example, in the case $b(\cS)=\cS$ and specifically in the regime~\eqref{def:hyp}, we have that $\Ub_\cS^0(\x(t))=0 \, \forall \x$, while in the regime~\eqref{def:hyp_v} in general $\Ub_\cS^0(\x(t))$ does not vanish and $\dot{\x}_i(t)=0$ if $\exists \x_i(t) \, : \, \Ub_\cS(\x_i(t))=0$. 
This does not tell us the position of the maxima, but as it is satisfied by all points, then it will be also satisfied by the maxima. 
In order to find the trajectory of the maxima and their position, we should look for the points $\bar{\x}_i$ s.t. 
\begin{equation}\label{min1}
    \varphi(t,\bar{\x}_i)=0
\end{equation}
that are the points of minimum of $\varphi$, i.e. 
\begin{equation}\label{min2}
    \nabla_\x \varphi(t,\bar{\x}_i)=0
\end{equation}
and
\begin{equation*}
D^2_\x \varphi(t,\bar{\x}_i)\ge 0.    
\end{equation*}
 Then, differentiating~\eqref{min1} with respect to time along the trajectories we find
\[
0=\dfrac{d}{dt}  \varphi(t,\bar{\x}_i)=\partial_t \varphi(t,\bar{\x}_i)+\dot{\bar{\x}}_i \cdot\nabla_\x \varphi(t,\bar{\x}_i).
\]
Because of~\eqref{min2}, then we have that
\[
\partial_t \varphi(t,\bar{\x}_i)=0
\]
and, therefore, from~\eqref{eq:HJ}
the minima satisfy $H(\bar{\x}_i,0)=0$.
Then we compute
\[
0=\dfrac{d}{dt} \nabla_\x \varphi(t,\bar{\x}_i)=\partial_t \nabla_\x \varphi(t,\bar{\x}_i)+\dot{\bar{\x}}_i D^2_\x \varphi(t,\bar{\x}_i).
\]
Differentiating~\eqref{eq:HJ} with respect to $\x$, we get 
\[
\partial_t \nabla_\x \varphi(t,\bar{\x}_i)=-\nabla_\x H(\x, \nabla_\x \varphi(t,\x)) - \nabla_p H(\x, \nabla_\x \varphi(t,\x)) D^2_\x \varphi(t,\x),
\]
and, specializing it in $\bar{\x}_i$, and remembering~\eqref{Hp0} and $H(\x,0)=0$, we get 
\[
\partial_t \nabla_\x \varphi(t,\bar{\x}_i)=- \nabla_p H(\x, \nabla_\x \varphi(t,\bar{\x}_i)) \cdot D^2_\x \varphi(t,\bar{\x}_i) =- \Ub_\cS^0(\bar{\x}_i(t))\cdot D^2_\x \varphi(t,\bar{\x}_i).
\]
Therefore we obtain 
\begin{equation*}
\dot{\bar{\x}}_i(t)=\Ub_\cS^0(\bar{\x}_i(t)),
\end{equation*}
that is the same as~\eqref{eq:traj}. In particular the  long term limit is solely determined by  $\Ub_\cS^0$ and does not depend on the initial condition of $\rho$ as it is usual in adaptive dynamics and as it follows from the H-Theorem in kinetic theory, that establishes in this linear case that the equilibrium is asymptotically stable and does not depend on the initial condition.
From the eikonal equation we may expect that
\[
\rho_0=\sum \dfrac{\rho^\infty}{\textrm{tr}\mathbb{D}_\cS^0(\bar{\x}_i)} \delta (x-\bar{\x}_i), \qquad \dot{\bar{x}}_i=\Ub_\cS(\bar{\x}_i)=0.
\]

\subsection{Examples}
We illustrate the results with two examples in one dimension. We first choose the signal
\begin{equation}\label{S1D}
    \cS(\x)=\cS_0\exp^{-\dfrac{(\x-\bar{\x})^2}{2\sigma^2}},
\end{equation}
with $\bar{\x}$ a given point in $\Omega$. When $b(\cS)=\cS$, it generates a transition probability given by
\begin{equation*}
    T[\cS](v,\hv)=\psi(v|\hv)\dfrac{\exp^{-\dfrac{R(\x-\bar{\x})\cdot \hv}{2\sigma^2}}}{\int_{\mathbb{S}^{d-1}}\exp^{-\dfrac{R(\x-\bar{\x})\cdot \hv}{2\sigma^2}} \, d\hv}.
\end{equation*}
Firstly, we remark that 
\begin{equation*}
    l_\cS=\dfrac{\sigma^2}{\max |\x-\bar{\x}|}.
\end{equation*}
In the regime~\eqref{def:hyp_v}, we expect that the singular point $\Ub_\cS^0(\x(t))=0$ only occurs when $\x(t)=\bar{\x}$. Moreover, we expect a unique nonhomogeneous stationary state whose profile needs to satisfy~\eqref{eq:infty}. In the regime~\eqref{def:hyp}, conversely, as the limiting $T[\cS]_0$ does not depend on $\x$, we expect that the homogeneous configuration is the stationary equilibrium, as $\partial_t\rho=0$.  
We now consider the Hopf-Cole analysis.
In 1D we have that $\hv=\pm 1$ and choosing $\psi(v|\hv)=\delta(v -\hv V)$ with $V,R=\mathcal{O}(1)$ we get
\begin{equation}\label{def:HsDs}
H(\x,p)=\dfrac{V^2p^2+Vp \mathcal{D}_\cS(\x)}{1+\sqrt{1+4V^2p^2+4Vp \mathcal{D}_\cS(\x)}}, \qquad 
\mathcal{D}_\cS(\x)=\dfrac{\exp^{\dfrac{R(\x-\bar{\x})}{2\sigma^2}}-\exp^{-\dfrac{R(\x-\bar{\x})}{2\sigma^2}}}{\exp^{\dfrac{R(\x-\bar{\x})}{2\sigma^2}}+\exp^{-\dfrac{R(\x-\bar{\x})}{2\sigma^2}}}.    \end{equation}
We remark that, therefore, in the regime \eqref{def:hyp_v}, the Hamiltonian has two different zeros $p(x)=0, -\dfrac{\mathcal{D}_\cS(\x)}{2V}$. Conversely, in the regime \eqref{def:hyp}, in~\eqref{def:HopfCole} (and, then, in~\eqref{def:HsDs}) $R$ is to be replaced with $\varepsilon R$ and, hence, the hamiltonian vanishes only in $p=0$ and every concentration disappears  as $\nabla_p H(x,0) \equiv 0$.

\begin{figure}[!htbp]
    \centering
\includegraphics[width=0.3\textwidth]{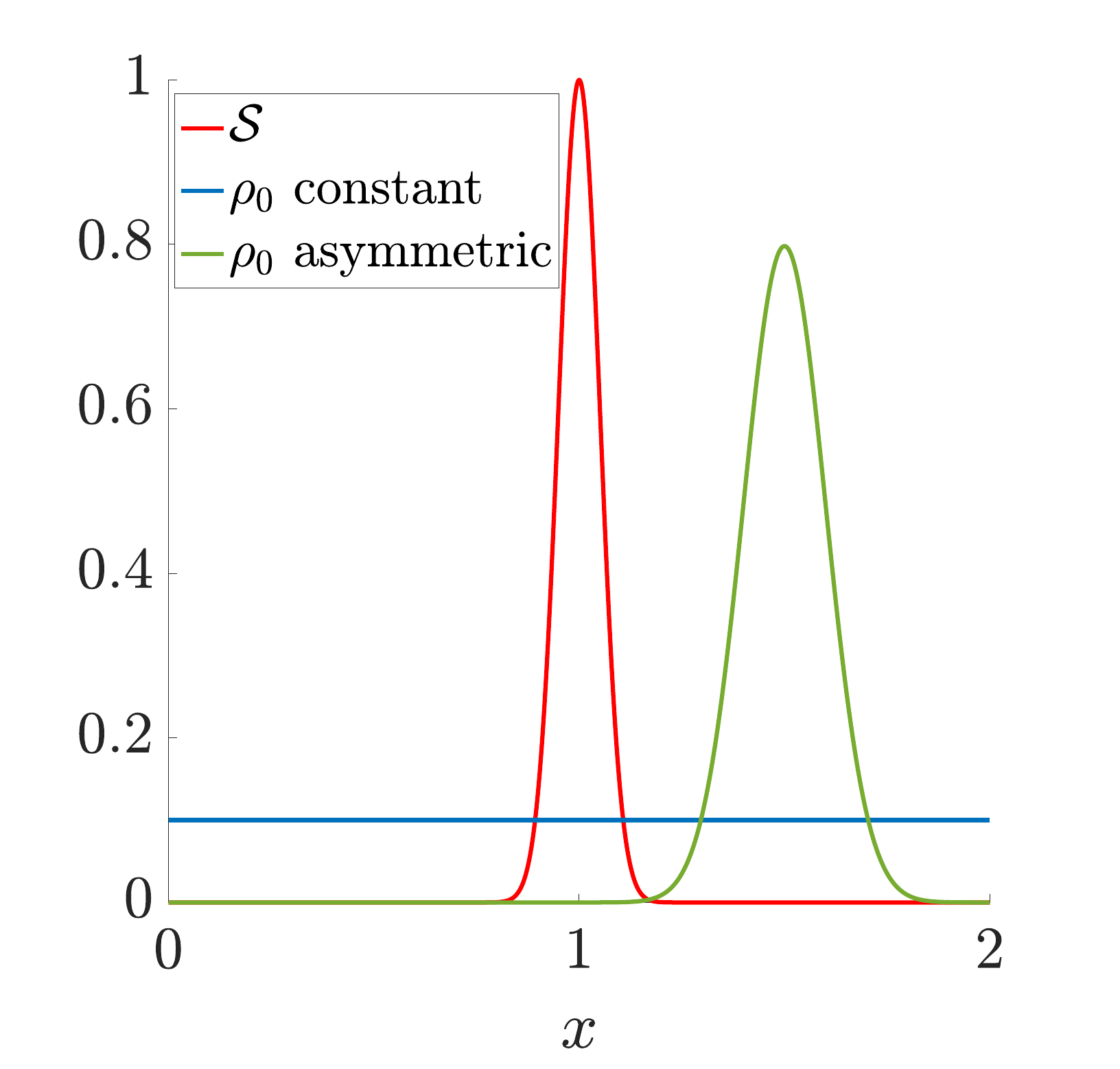}
\includegraphics[width=0.3\textwidth]{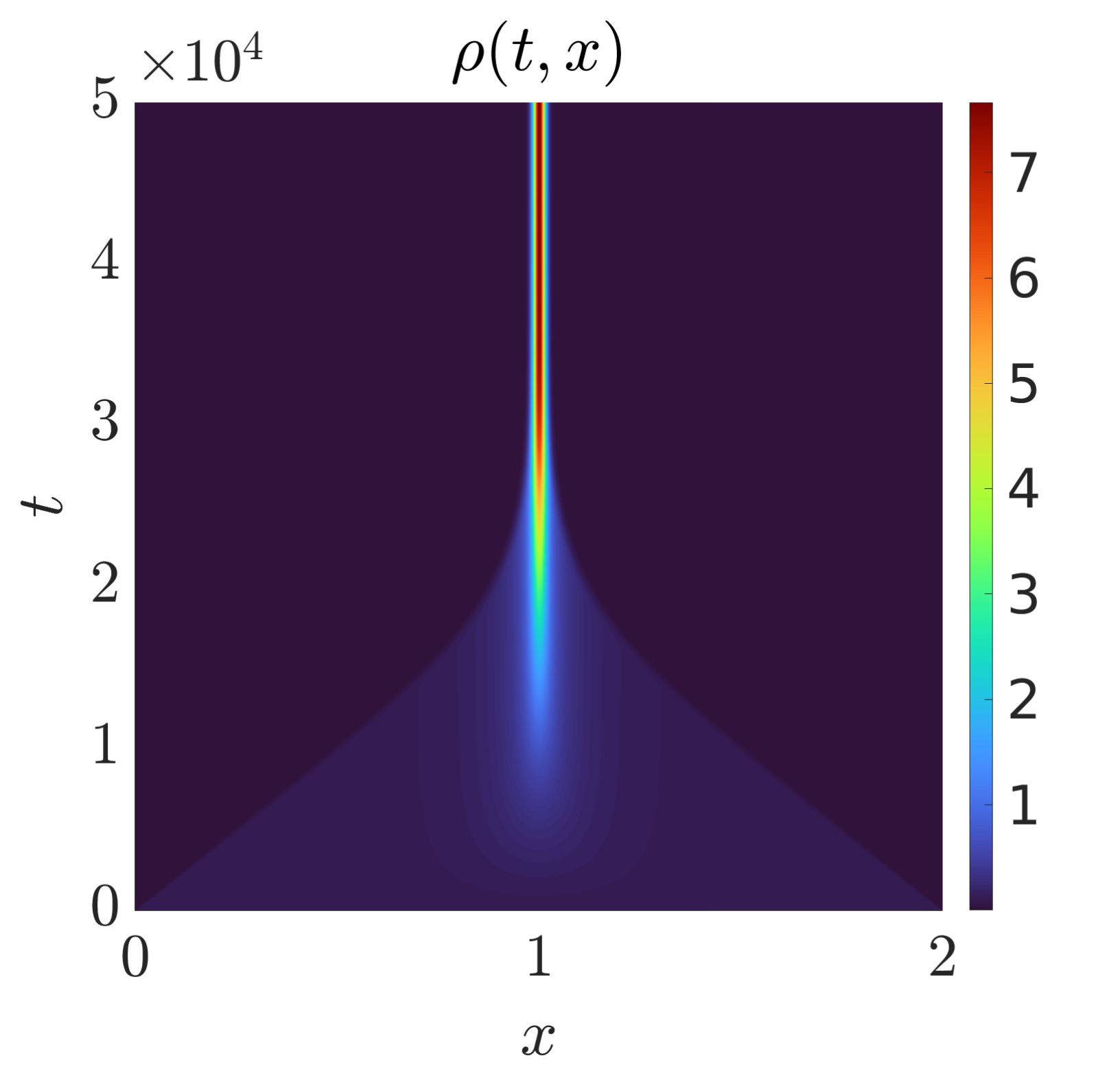}
\includegraphics[width=0.3\textwidth]{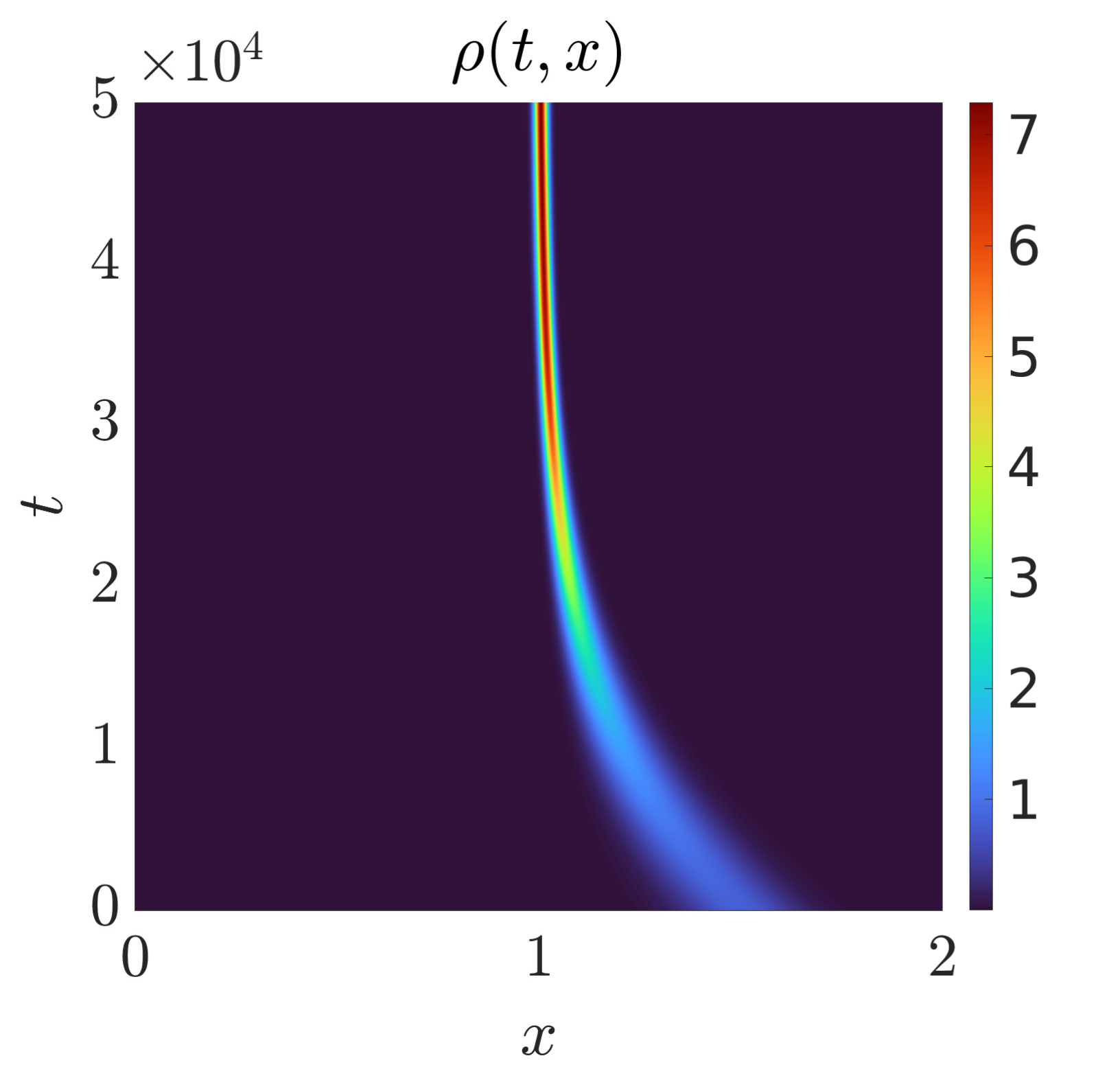}\\
\includegraphics[width=0.33\textwidth]{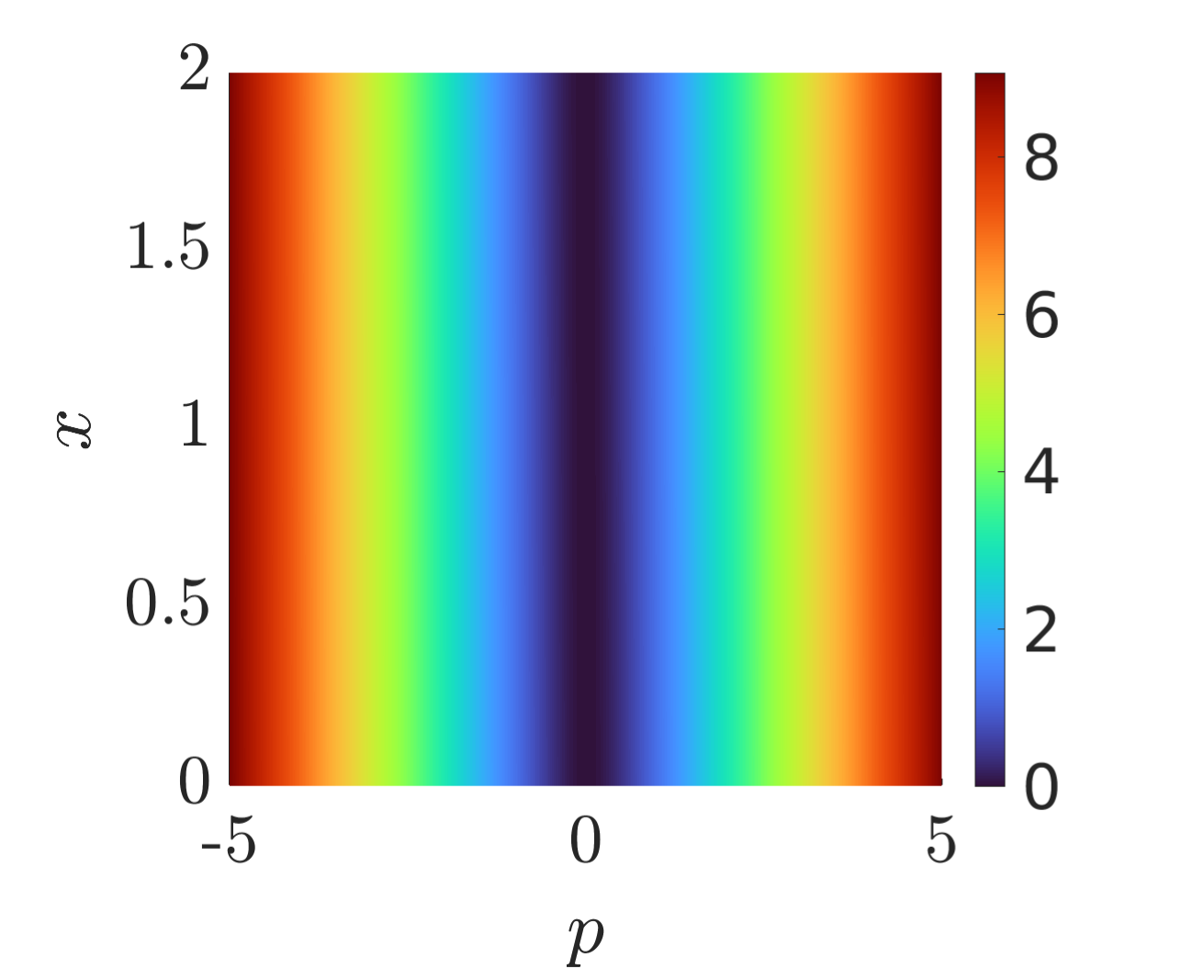}
\hspace{-0.3cm}
\includegraphics[width=0.33\textwidth]{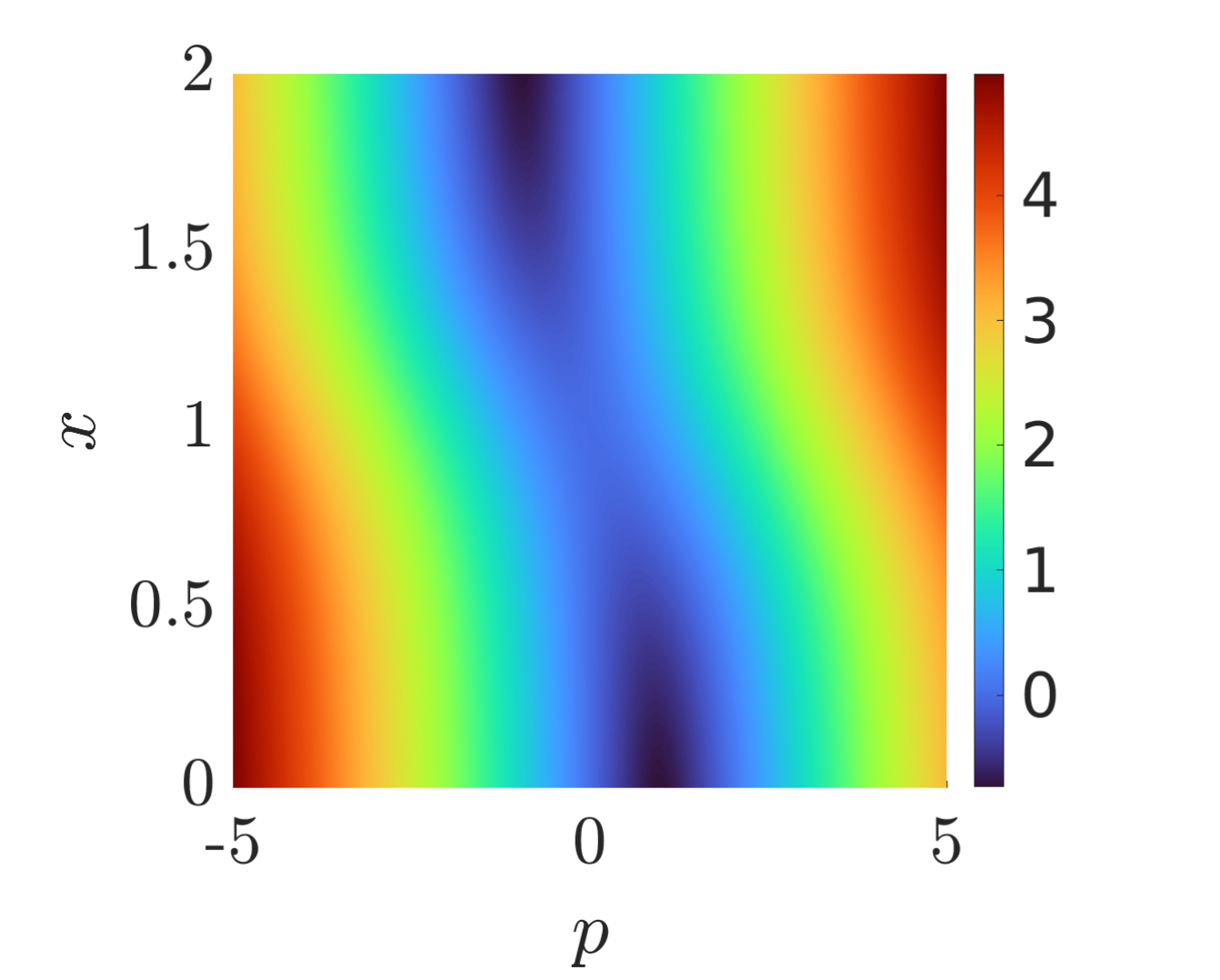}
\hspace{-0.4cm}
\includegraphics[width=0.33\textwidth]{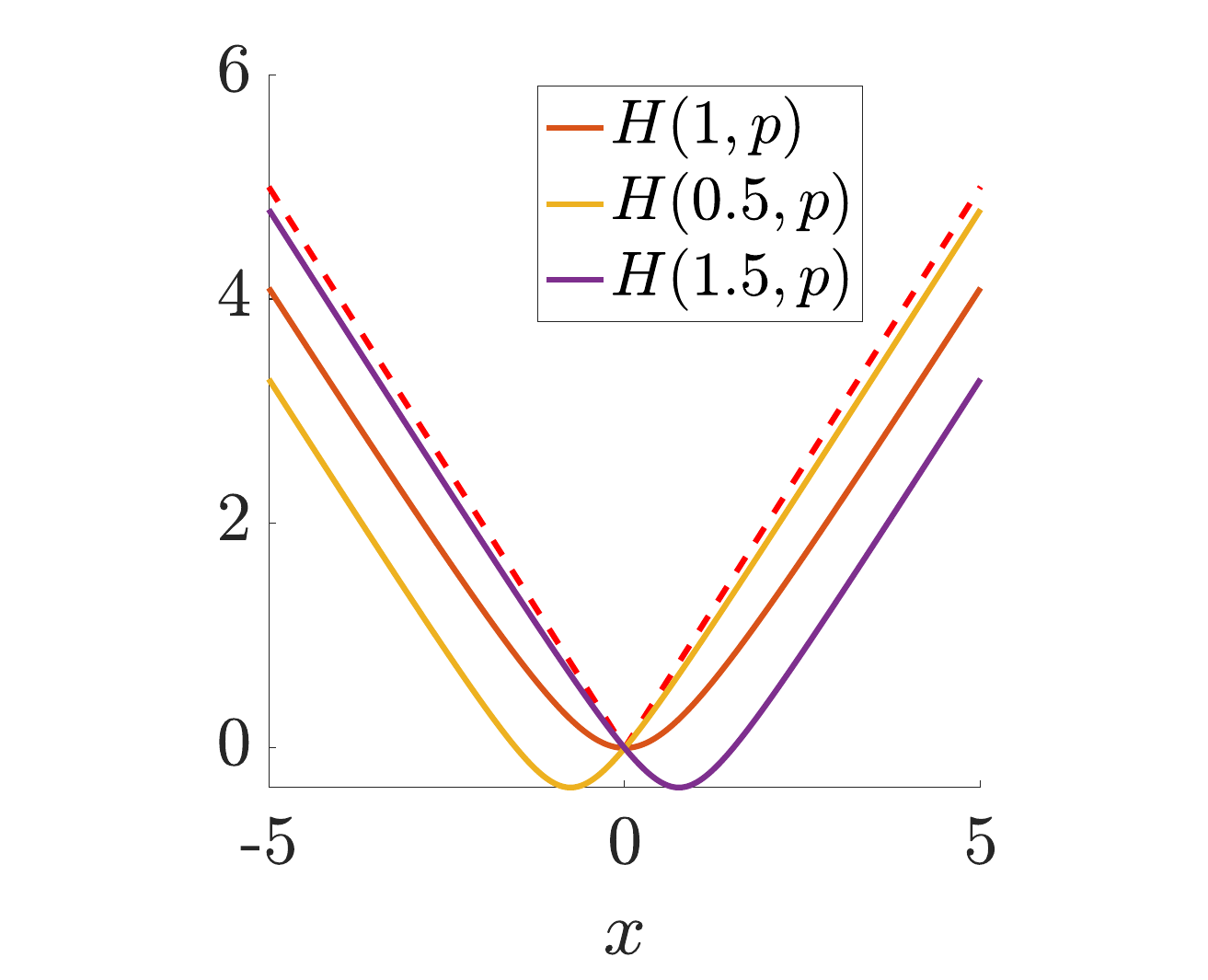}
     \vspace{-10pt}
    \caption{First line. First panel: $\cS$ (red curve) given by~\eqref{S1D} and two different initial conditions $\rho^0$: the constant one (blue) and an asymmetric Gaussian (green). Second panel: temporal evolution of $\rho(t,\x)$ in the case of constant $\rho^0$. Third panel: temporal evolution of $\rho(t,\x)$ in the case of an asymmetric $\rho^0$.  Second line. First panel: $H(\x,p)$ in the local regime~\eqref{def:hyp}, second and third panel: $H(\x,p)$ in the nonlocal regime~\eqref{def:hyp_v}. In the third panel the red dashed lines correspond to $\pm U|p|$, while the black horizontal dashed line indicates the level zero.}
    \label{fig1}
\end{figure}

We numerically solve equation~\eqref{eq:cinetique}.
We consider $\Omega=[0,1]$, $\mu=1$ and $V=5\cdot 10^{-5}$ and $\psi(v)=\dfrac{1}{V}$. In all simulations the space grid has a uniform mesh defined by $dx=10^{-3}$.
In Fig.~\ref{fig1} we use $\cS$ given by~\eqref{S1D} with $\sigma=0.05$ and $\bar{\x}=1$, $R=0.01$. Therefore $l_\cS=5\cdot 10^{-3}$. Let us consider $L=1$. We are then in regime~\eqref{def:hyp_v} with $\varepsilon=10^{-5}$. We consider two different initial conditions $\rho^0=0.1$ and $\rho^0$ Gaussian centered in $1.5$. As already mentioned the stationary state is unique and does not depend on the initial condition. In the second line of Fig~\ref{fig1}, second and third panel, we plot the Hamiltonian \eqref{def:HsDs}. We remark that the Hamiltonian is not always positive and there is a concentration profile.
Conversely, in a regime defined by $V=5\cdot 10^{-5}$, $L=l_\cS$ (i.e. regime~\eqref{def:hyp} with $\varepsilon=10^{-2}$) the stationary state $\rho_{\infty}$ is the stationary homogeneous configuration even for a nonhomogeneous initial condition (not shown), and this is true in both regimes $R\lessgtr l_\cS$. The corresponding Hamiltonian is plotted in Fig.\ref{fig1}, second line, first panel. 

Second, we choose a bimodal signal
\begin{equation}\label{S2D}
    \cS(\x)=\cS_1\exp^{-\dfrac{(\x-\bar{\x}_1)^2}{2\sigma_1^2}}+\cS_2\exp^{-\dfrac{(\x-\bar{\x}_2)^2}{2\sigma_2^2}}.
\end{equation}
Then the number of singular points $\x_i$ satisfying $\Ub_\cS(\x_i)=0$  depends on $R$, $\bar{\x}_1-\bar{\x}_2$ and on $\sigma^2$.

\begin{figure}[!htbp]
    \centering
    \subfigure[$\cS$]{\includegraphics[width=0.5\textwidth]{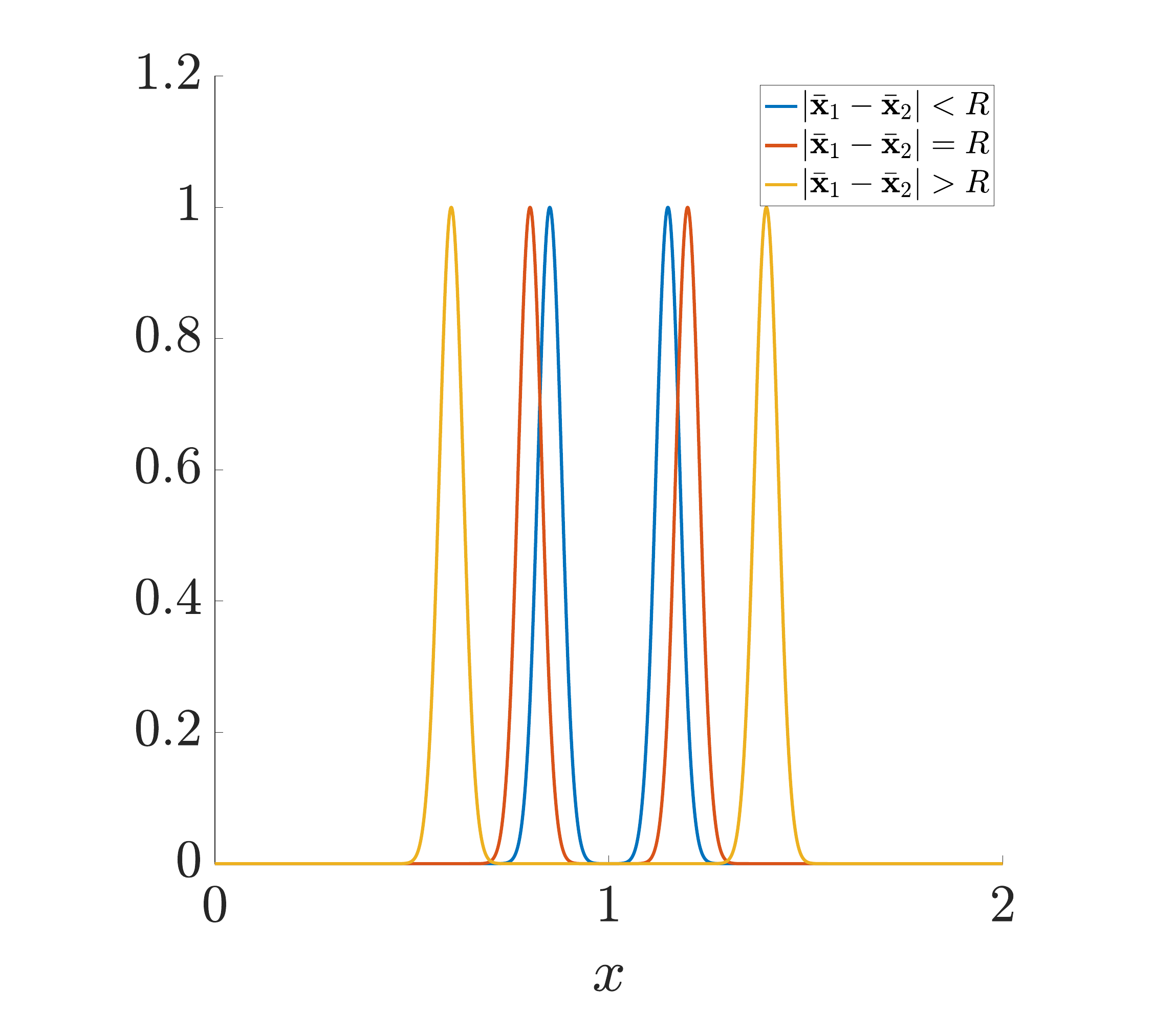}}
    \hspace*{-0.7cm}
    \subfigure[$\bar{\x}_2-\bar{\x}_1 =R$]{\includegraphics[width=0.45\textwidth]{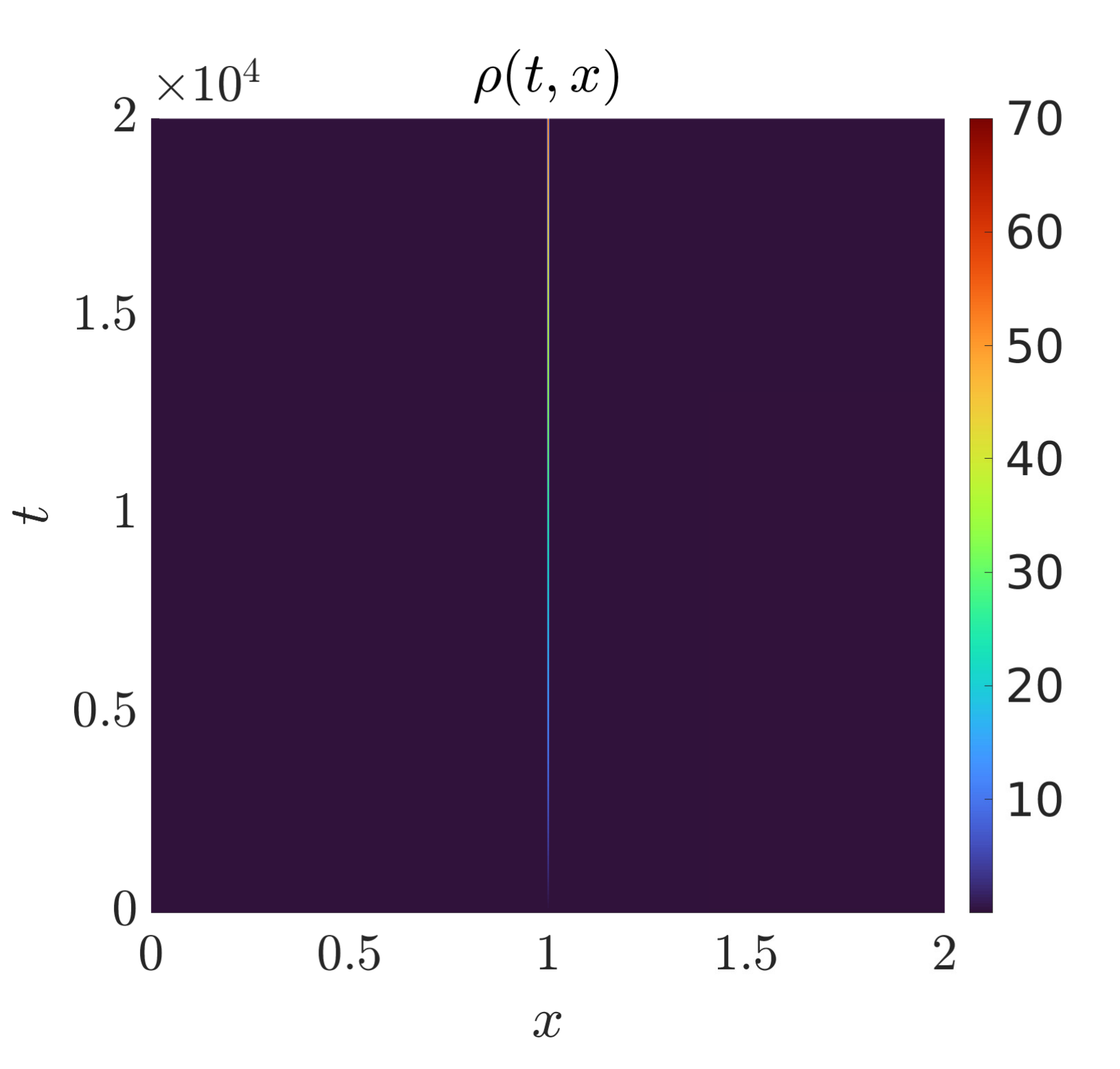}}\\
    \subfigure[$\bar{\x}_2-\bar{\x}_1 >R$]{\includegraphics[width=0.45\textwidth]{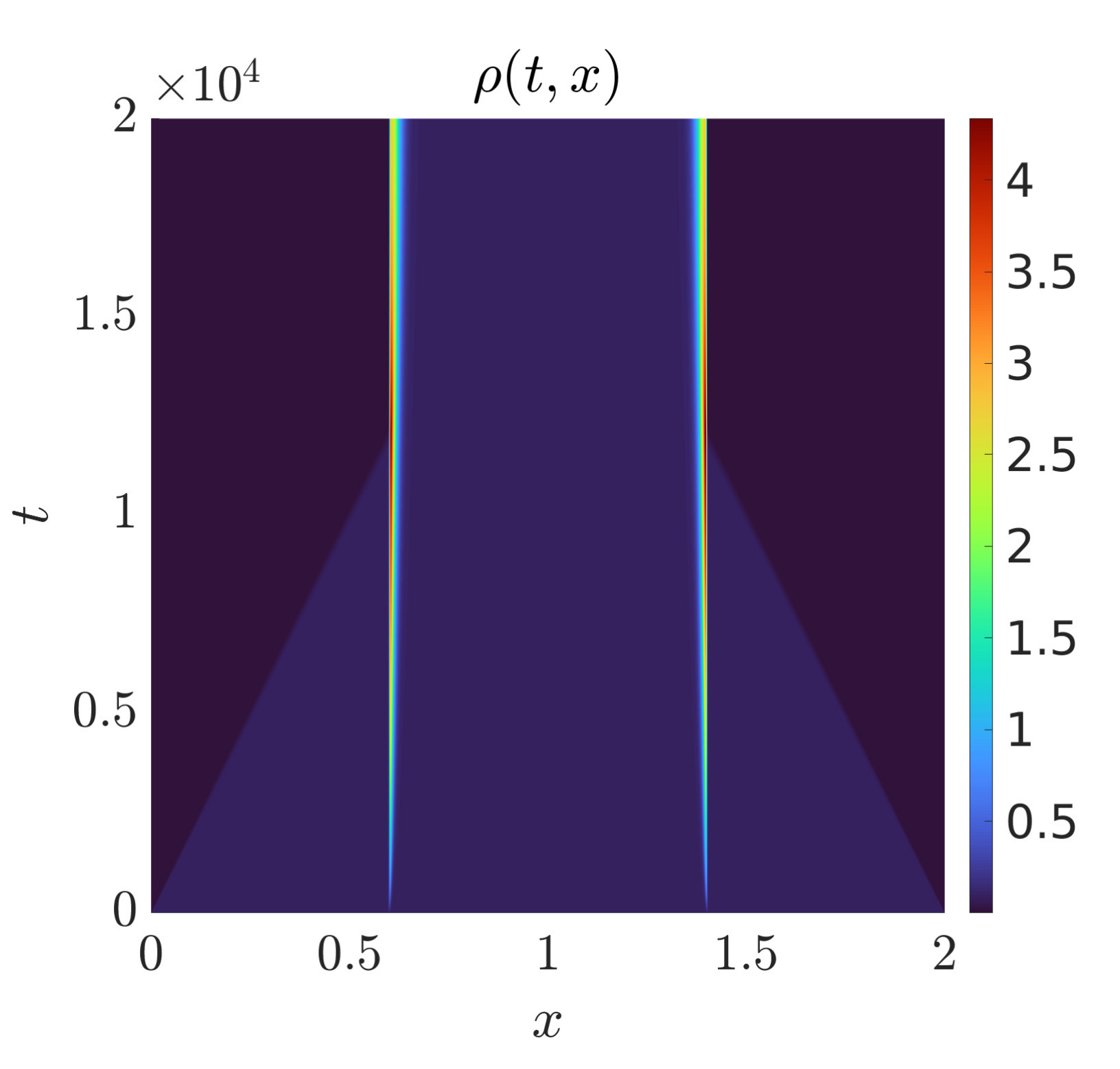}}
    \subfigure[$\bar{\x}_2-\bar{\x}_1 >R$]{\includegraphics[width=0.45\textwidth]{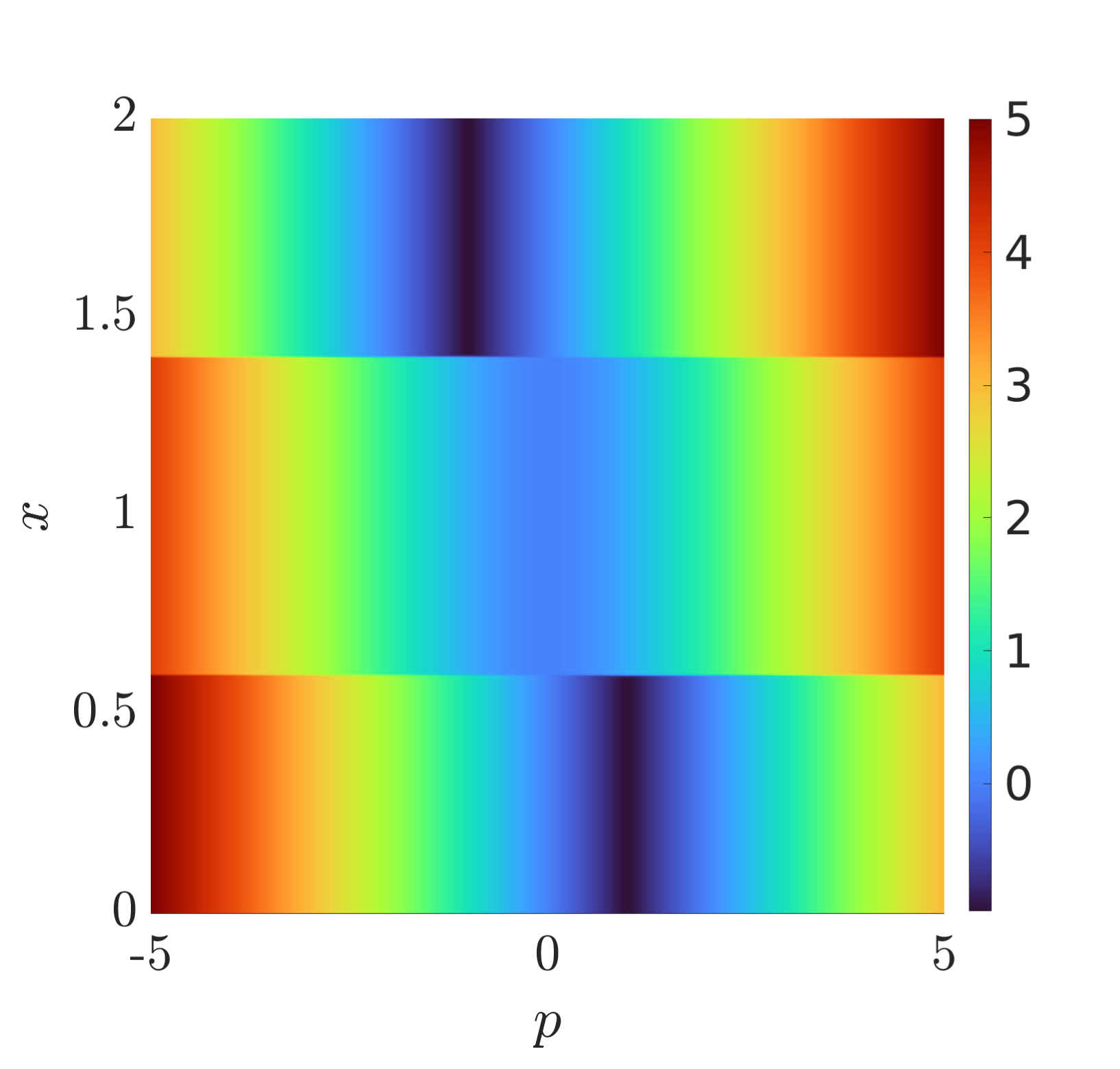}}
     \vspace{-10pt}
    \caption{Temporal evolution of $\rho(t,\x)$ in case of $\cS$ given by~\eqref{S2D}. Here $R=0.4$, the initial condition $\rho^0$ is constant. In (a): three different $\cS$ as given by~\eqref{S2D} with $\sigma_1=\sigma_2=0.03$ and three different couples of $\bar{\x}_2, \bar{\x}_1$. In (b)-(c) temporal evolution of $\rho(t,\x)$ for the two different $\cS$: in (b) $\bar{\x}_2-\bar{\x}_1=R$, in (c) $\bar{\x}_2-\bar{\x}_1>R$. In (d) we plot the Hamiltonian for the case $\bar{\x}_2-\bar{\x}_1>R$.}
    \label{fig2}
\end{figure}

In Fig.~\ref{fig2} we consider the bimodal signal $\cS$ given by~\eqref{S2D} with $\sigma_1=\sigma_2=0.03$ and three different couples of $\bar{\x}_1, \bar{\x}_2$ according to the value of their distance with respect to $R=0.4$. Here again $\varepsilon=10^{-5}$ (same values of $V,L, \mu$). We remark that, when $\bar{\x}_2-\bar{\x}_1 \le R$ (see Fig.~\ref{fig2}(b) for the case $\bar{\x}_2-\bar{\x}_1=R$), then there is a single peak as $\exists !$ $\x_1(=0)$ such that $\Ub_\cS(\x_1)=0$. The case $\bar{\x}_2-\bar{\x}_1<R$ behaves like $\bar{\x}_2-\bar{\x}_1=R$ (not shown). When $\bar{\x}_2-\bar{\x}_1 >R$ (see Fig.~\ref{fig2}(c)) then $\exists$ $\bar{\x}_1, \bar{\x}_2$ such that $\Ub_\cS(\bar{\x}_i)=0, \, i=1,2$. In Fig.~\ref{fig2}(d) we plot the Hamiltonian in the case $\bar{\x}_2-\bar{\x}_1 >R$.

\begin{figure}[!htbp]
    \centering
    \subfigure[$\cS$]{\includegraphics[width=0.3\textwidth]{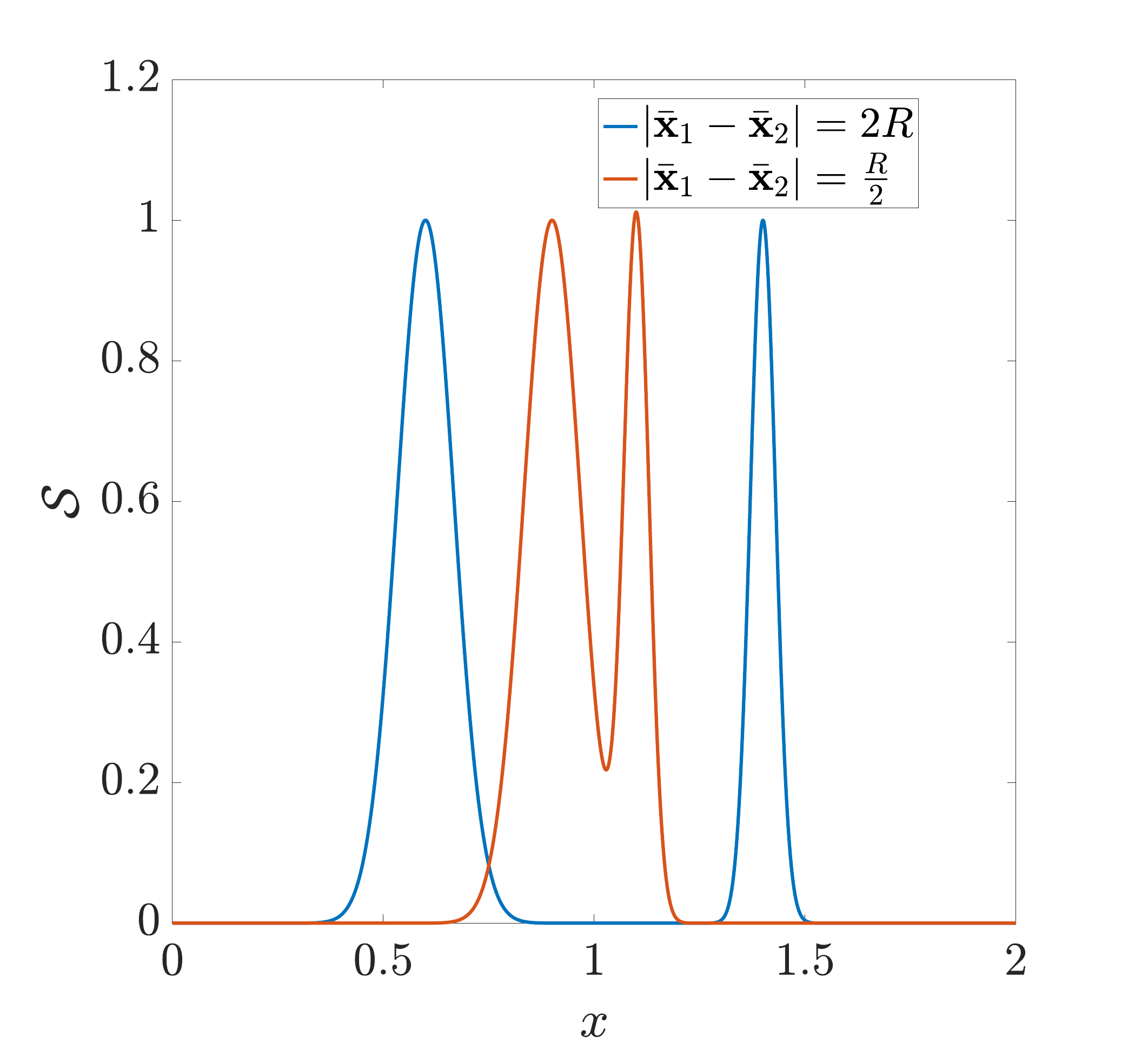}}
    \subfigure[$\bar{\x}_2-\bar{\x}_1=2R$]{\includegraphics[width=0.32\textwidth]{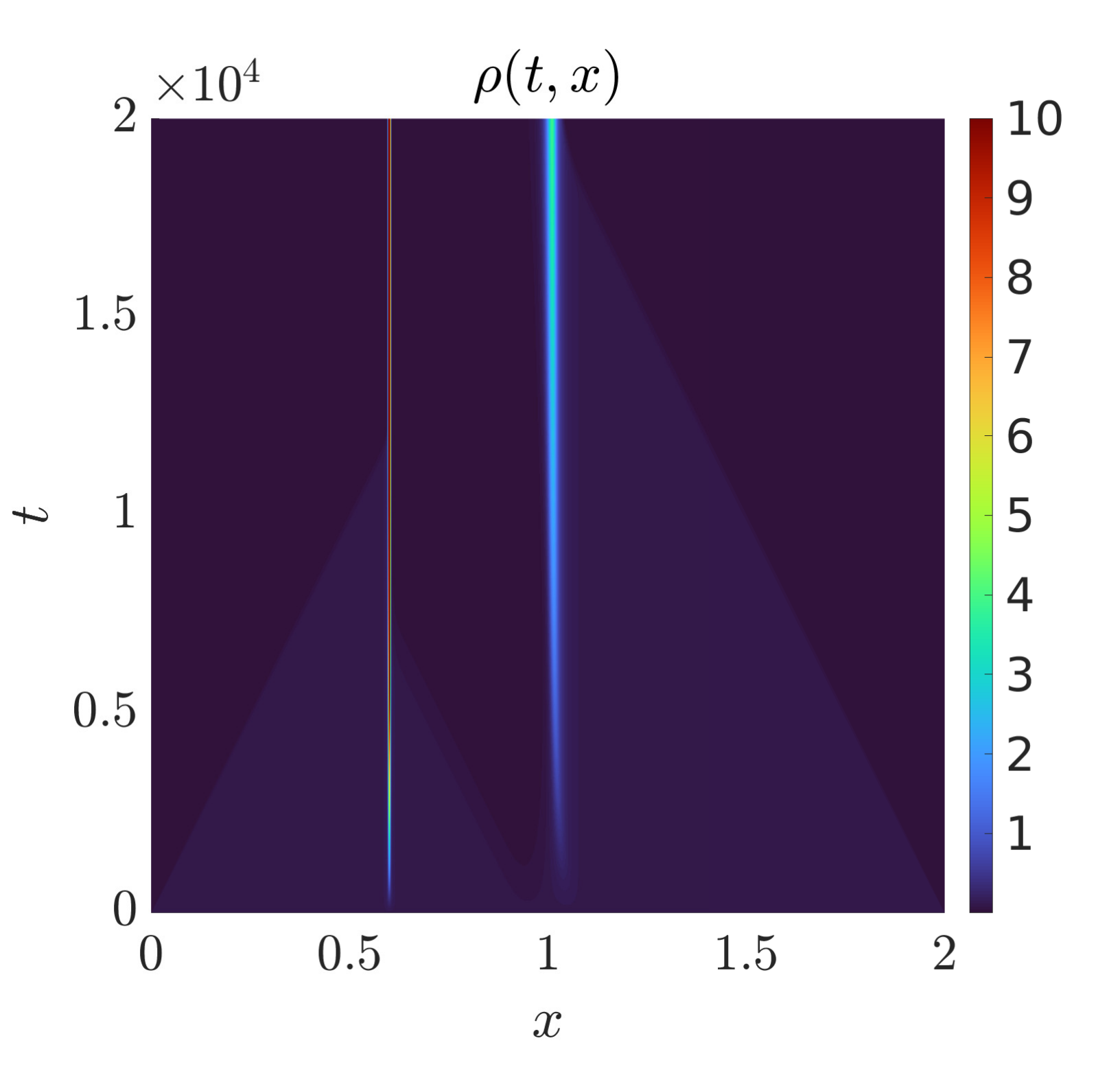}}
    \subfigure[$\bar{\x}_2-\bar{\x}_1=R/2$]{\includegraphics[width=0.32\textwidth]{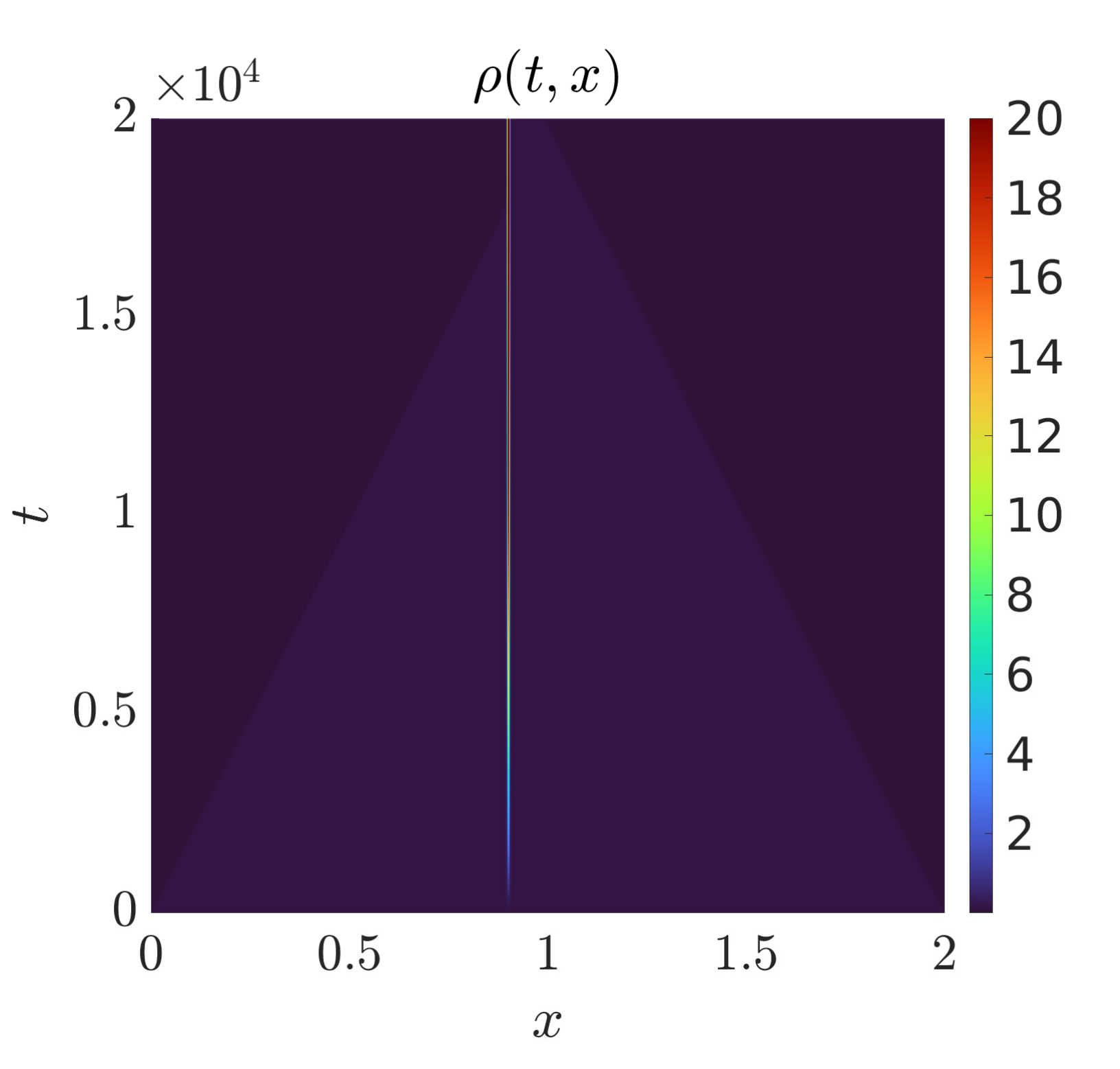}}
     \vspace{-10pt}
    \caption{Temporal evolution of $\rho(t,\x)$ in case of $\cS$ given by~\eqref{S2D}. Here $R=0.4$ and $\sigma_1=0.06, \sigma_2=0.03$. In (a): two different $\cS$ given by~\eqref{S2D} with two different couples $\bar{\x}_2, \bar{\x}_1$. In (b) and (c) the corresponding time evolution of the densities $\rho$.}
    \label{fig3}
\end{figure}

In Fig. \ref{fig3} we consider $\cS$ given by~\eqref{S2D} with $\sigma_1=2\sigma_2, \sigma_2=0.03$ while for the sensing radius we have again $R=0.4$. We consider two different couples of $\bar{\x}_1,\bar{\x}_2$ as shown in Fig. \ref{fig3}(a). We remark that the peaks of $\rho$ do not coincide with the peaks of $\cS$ and this is due to the nonlocality ($R>0$). In particular, in the first case (b) the distance between the maxima of $\cS$ is larger than the sensing radius and, thus, the stationary solution has two peaks even though with different convexity due to the configuration of $\cS$. In the second case (c), the distance between the peaks of $\cS$ is exactly $R/2$, so that the peak of the stationary solution is unique, even though it is asymmetric, because of the asymmetry of $\cS$.

In conclusion, in this linear case the analysis of the kinetic equations and of the aggregate limits give almost a complete set of information concerning the dynamics of the maxima, except the concentration result, for which the WKB analysis is needed. Therefore, we now consider a nonlinear case in which the study of the kinetic and aggregate equations may not be able to convey all the necessary information regarding the dynamics of the maxima points, while the Hamilton-Jacobi formalism offers promising tools in order to describe the concentration profiles.



\section{A nonlinear case}

When the external field affecting the choice of the reorientation is the cell density itself, i.e., $\cS=\rho$, then Eq.~\eqref{eq:cinetique} becomes nonlinear. The kinetic equation is
\begin{equation}\label{eq:cinetique_adh}
    \partial_t f(t,\x,v,\hv) +\vb \cdot \nabla_\x f(t,\x,v,\hv)= \mu \left(\rho(t,\x)T[\rho](v,\hv)-f(t,\x,v,\hv)\right),
\end{equation}
where 
\begin{equation}\label{def:Trho}
    T[\rho]=\dfrac{\rho(t,\x+R\hv)}{\int_{\mathbb{S}^{d-1}}\rho(t,\x+ R\hv)\, d\hv}\psi(v|\hv).
\end{equation}
In \cite{loy2020KRM} the authors perform a linear stability analysis. In 1D and with the choice $\psi(v|\hv)=\delta(v-V_\psi(\hv))$, they show that the uniform homogeneous configuration is stable if, using the notation~\eqref{def:Vpsi}, 
\begin{equation}\label{eq:stab}
    \dfrac{V}{R \mu}>1, \qquad V=\dfrac{V_\psi(+1)+V_\psi(-1)}{2}.
\end{equation}

We now consider regime~\eqref{def:hyp}, and in this  nondimensionalized regime $V,R,\mu=\mathcal{O}(1)$, then Eq.~\eqref{eq:cinetique_adh} reads
\begin{equation}\label{eq:2}
  \dfrac{\partial f_{\varepsilon}}{\partial t}(t,\x,v,\hv) + \vb\cdot \nabla f_{\varepsilon}(t,\x,v,\hv) =  \dfrac{\mu}{\varepsilon}  \, \Big( \rho_\varepsilon T[\rho]_\varepsilon - f_{\varepsilon}(t,\x,v,\hv) \Big),
\end{equation}
where
\[
T[\rho]_\varepsilon=\dfrac{\rho_{\varepsilon}(t,\x+\varepsilon R\hv)}{\int_{\mathbb{S}^{d-1}}\rho_{\varepsilon}(t,\x+\varepsilon R\hv)\, d\hv}\psi(v|\hv).
\]
In particular, in the rescaled regime~\eqref{def:hyp}, relation~\eqref{eq:stab} is unchanged since
\[
\dfrac{\varepsilon V}{R\varepsilon \mu}=\dfrac{V}{R \mu}.
\]

At the macroscopic level we have that
\begin{equation}\label{eq:macro_rho}
    \partial_t \rho_\varepsilon +\nabla_\x \cdot (\Ub_\rho^\varepsilon \rho_\varepsilon)=\varepsilon \nabla_\x\cdot \nabla_\x\cdot \left( \mathbb{D}_\rho^\varepsilon \rho\right)
\end{equation}
with $\Ub_\rho^\varepsilon$ and $\mathbb{D}_\rho^\varepsilon$ defined thanks to~\eqref{defU},~\eqref{defD}.

When $\rho_{\varepsilon}$ is smooth enough, the limiting transition probability becomes
\[
\lim_{\varepsilon \rightarrow 0} T[\rho]_\varepsilon = T[\rho]_0:=\dfrac{1}{|\mathbb{S}^{d-1}|}\psi(v|\hv).
\]

\subsection{Concentration profile and the Hamilton–Jacobi equation}
However, when $\rho_{\varepsilon}$ undergoes concentration, we may consider~\eqref{def:HopfCole} and we have
\begin{align*}
\begin{aligned}[b]
    &T[\rho]_\varepsilon= \psi(v|\hv)\dfrac{\int_0^U \int_{\mathbb{S}^{d-1}}\exp^{-\dfrac{\varphi_\varepsilon(t,\x+\varepsilon R\hv,w,\hw)}{\varepsilon}} dw d\hw}{\int_{\mathbb{S}^{d-1}}\int_0^U \int_{\mathbb{S}^{d-1}}\exp^{-\dfrac{\varphi_\varepsilon(t,\x+\varepsilon R\hv,w,\hw)}{\varepsilon}} \, dw d\hw d\hv}\\
    &\approx  \psi(v|\hv)\dfrac{\int_0^U \int_{\mathbb{S}^{d-1}}\exp^{\dfrac{-\varphi_\varepsilon(t,\x,w,\hw)-\varepsilon R\hv\cdot \nabla \varphi_\varepsilon(t,\x,w,\hw)}{\varepsilon}} \, dw d\hw}{\int_0^U \int_{\mathbb{S}^{d-1}}\int_{\mathbb{S}^{d-1}}\exp^{-\dfrac{\varphi_\varepsilon(t,\x,w,\hw)+\varepsilon R\hv \cdot\nabla_\x  \varphi}{\varepsilon}}   dw d\hw \, d \hv} 
\end{aligned}
\end{align*}
where we have used
\[
\varphi_\varepsilon(t,\x+\varepsilon R\hv,w,\hw)=\varphi_\varepsilon(t,\x,w,\hw)+\varepsilon R\hv\cdot \nabla_\x \varphi_\varepsilon(t,\x,w,\hw).
\]
Then, assuming~\eqref{defQphieps} and, remembering that $\tilde{\varphi}_\varepsilon$ does not depend on $v,\hv$,  therefore
\[
T[\rho]_\varepsilon \approx  \psi(v|\hv) \dfrac{\exp^{\dfrac{- \tilde \varphi_\varepsilon(t,\x)}{\varepsilon}}
\exp^{- R\hv\cdot \nabla \tilde \varphi_\varepsilon(t,\x)}\int_0^U 
\int_{\mathbb{S}^{d-1}}Q_\varepsilon \, dw d\hw}{\exp^{\dfrac{- \tilde \varphi_\varepsilon(t,\x)}{\varepsilon}}\int_{\mathbb{S}^{d-1}}\int_0^U \int_{\mathbb{S}^{d-1}}Q_\varepsilon dw d\hw \exp^{- R \hv \cdot \nabla \tilde \varphi_\varepsilon}    \, d \hv},
\]
where we have used the Fubini-Tonelli theorem.
Then, letting $\varepsilon \rightarrow 0^+$, we obtain
\begin{equation}\label{def:G}
    T[\rho]_\varepsilon \rightarrow G_R(v,\hv,\nabla_\x \varphi)= \psi(v|\hv) \dfrac{\exp^{- R\hv \cdot \nabla_\x \varphi}}{\displaystyle\int_{\mathbb{S}^{d-1}} \exp^{- R\hv \cdot \nabla_\x \varphi} \, d\hv}.
\end{equation}

Therefore, plugging~\eqref{def:HopfCole} in~\eqref{eq:2} we obtain
\begin{equation}\label{eq:tr.1}
\begin{aligned}[b]
    \mu-\partial_t \varphi_\varepsilon-\vb \cdot \nabla_\x \varphi_\varepsilon =& \exp^{\dfrac{\varphi_\varepsilon(t,\x,v,\hv)}{\varepsilon}}
    \\
    &\dfrac{\mu}{\varepsilon}\left[\int_0^U \int_{\mathbb{S}^{d-1}}\exp^{-\dfrac{\varphi_\varepsilon(t,\x,w,\hw)}{\varepsilon}}\!\!\!\!  dw d\hw  \; 
    T[\rho]_\varepsilon  \right] .
\end{aligned}
\end{equation}
Furthermore, considering the expansion~\eqref{defQphieps} and by letting (formally) $\varepsilon$ go to $0^+$, we obtain
\begin{equation}\label{eq:tr.2}
  \begin{aligned}[b]
\mu-\partial_t \varphi &-\vb \cdot \nabla_\x \varphi\\
    &=\mu\left[Q^{-1}(\x,v,\hv)    \int_0^U \int_{\mathbb{S}^{d-1}} Q(\x,w,\hw) \, dw d\hw  G_R(v,\hv,\nabla_\x \varphi) \right].
    \end{aligned}
\end{equation}
Again, we assume that $Q(\x,w,\hw)$ is positive according to assumption~\eqref{as:H}. 
Then, like in the previous section, we define an eigenvalue-eigenvector problem
\begin{equation*}\label{pb:eigenvalue}
  \begin{aligned}[b]
    &(\mu+H(p)-v\hv\cdot p){\mathcal Q} (p,v,\hv)=\mu   G_R(v,\hv,p) \int_0^U \int_{\mathbb{S}^{d-1}} {\mathcal Q} (p,w,\hw)   \, dw d\hw.
    \end{aligned}
\end{equation*}
We remark that the term $G_R$ results from the interaction kernel and arises due to the nonlocal sensing of $\rho$. As such, its role is to drive the dynamics of $f$ toward the equilibrium in $v,\hv$. We remark that it satisfies: 
\[
\nabla_p G_R(v,\hv,p)= \psi(v|\hv)\left[
\dfrac{-R\hv\exp^{- R\hv \cdot p}}{\displaystyle\int_{\mathbb{S}^{d-1}} \exp^{- R\hw \cdot p} \, d\hw}
+\dfrac{\exp^{- R\hv \cdot p}R \displaystyle\int_{\mathbb{S}^{d-1}} \exp^{- R\hw \cdot p}\hw \, d\hw}{\left(\displaystyle\int_{\mathbb{S}^{d-1}} \exp^{- R\hw \cdot p} \, d\hw\right)^2} \right],
\]
and, then
$$
G_R(v,\hv,0)=\dfrac{\psi(v|\hv)}{|\mathbb{S}^{d-1}|}, \qquad \nabla_p G_R(v,\hv,0)=-\psi(v|\hv)\dfrac{R}{|\mathbb{S}^{d-1}|}\hv.
$$

From~\eqref{eq:tr.1},~\eqref{eq:tr.2}, we find that the (formal) limit $\varphi$ is the solution of
\begin{equation}\label{HJ_nonlinear}
\partial_t \varphi +H(\nabla_\x \varphi)=0,
\end{equation}
where the Hamiltonian is implicitly defined by
\begin{equation}\label{pb:eigenvalue1}
    1=\mu \int_0^U \int_{\mathbb{S}^{d-1}}\dfrac{ G_R(v, \hv,p)}{\mu +H(p)-v \hv \cdot p} \, dv d\hv.
\end{equation}
We remark that now $H$ only depends on $p$ and not on $\x$. It is easy to see that $H(0)=0$.
Then, by differentiating~\eqref{pb:eigenvalue1} with respect to $p$ we obtain
\[
0=\mu \int_0^U \int_{\mathbb{S}^{d-1}}\dfrac{ \nabla_p G_R(v,\hv,p)}{(\mu +H(p)-v \hv \cdot p)} \, dv d\hv-\mu \int_0^U \int_{\mathbb{S}^{d-1}}\dfrac{ G_R(v,\hv,p)(\nabla_p H(p)-v\hv)}{(\mu +H(p)-v \hv \cdot p)^2} \, dv d\hv.
\]
Therefore $\nabla_p H(p)=\Ub_{R}^G (p)$, where
\begin{equation} \label{eq:URG}
\Ub_R^G(p)=\dfrac{\int_{\mathbb{S}^{d-1}}V_\psi(\hv)\exp^{-R\hv \cdot \nabla_\x \varphi} \, \hv d\hv}{\int_{\mathbb{S}^{d-1}}V_\psi(\hv)\exp^{-R\hv \cdot \nabla_\x \varphi} \, d\hv},
\end{equation}
so that
\begin{equation}\label{eq:nablaH}
 \nabla_p H(0)=\int_{\mathbb{S}^{d-1}} V_\psi(\hv) \hv \, d\hv=\Ub_R^G(0).   
\end{equation}
We remark that $\nabla_p H(0)$ vanishes in the case in which $V_\psi$ is even.

Differentiating further, the Hessian of $H$ satisfies
\[
\begin{aligned}[b]
\int_0^U \int_{\mathbb{S}^{d-1}}&\dfrac{G_R(v,\hv,p) D^2 H(p)}{(\mu+H-v\hv p)^2} \, dv d\hv =2 \int_0^U \int_{\mathbb{S}^{d-1}}\dfrac{G_R (\nabla_p H-v \hv)\otimes (\nabla_p H-v \hv)}{(\mu+H-v\hv p)^3} \, dv d\hv\\
&+ \int_0^U \int_{\mathbb{S}^{d-1}}\dfrac{D_p^2 G_R}{\mu+H-v\hv p} \, dv d\hv-2\int_0^U \int_{\mathbb{S}^{d-1}}\dfrac{\nabla_p G_R (\nabla_p H-v \hv)}{(\mu+H-v\hv p)^2} \, dv d\hv.
\end{aligned}
\]
We have $D_p^2 G_R(v,\hv,0)=0$ because 
\begin{align*}
D_p^2 G_R(v,\hv,p)=&\psi(v|\hv)R^2\exp^{-R \hv \cdot p}\left[\dfrac{\hv \otimes \hv }{\int_{\mathbb{S}^{d-1}} \exp^{- R\hv \cdot p} \, d\hv} 
-2 \dfrac{\hv \otimes \int_{\mathbb{S}^{d-1}} \exp^{- R\hv \cdot p} \hv\, d\hv }{(\int_{\mathbb{S}^{d-1}} \exp^{- R\hv \cdot p} \, d\hv)^2}
\right.
\\
& \left.-\dfrac{\int_{\mathbb{S}^{d-1}} \exp^{- R\hv \cdot p}\hv \otimes \hv \, d\hv }{(\int_{\mathbb{S}^{d-1}} \exp^{- R\hv \cdot p} \, d\hv)^2} +2 \dfrac{\int_{\mathbb{S}^{d-1}} \exp^{- R\hv \cdot p} \, d\hv \otimes \int_{\mathbb{S}^{d-1}} \exp^{- R\hv \cdot p} \, d\hv }{(\int_{\mathbb{S}^{d-1}} \exp^{- R\hv \cdot p} \, d\hv)^3} \right].
\end{align*}
Then, in the case in which $\psi$ does not depend on $\hv$, we have $V=V_\psi$ and
\begin{align} \label{eq:D2H}
D_p^2 H(0) =\dfrac{2dD^2}{\mu}\mathbb{I} -2dV R \mathbb{I}= 2d \big(\dfrac{D^2}{\mu}-V R \big) \; \mathbb{I}
\end{align}
that is positive definite when 
\begin{equation}\label{eq:stab_H}
    \dfrac{D^2}{V}>R \mu.
\end{equation} 
Notice that $D^2=V^2$ when $\psi$ is  a Dirac delta and we recover, here in any dimension $d$, the linear stability region determined in 1D in~\cite{loy2020KRM}. For any other choice of $\psi$, we have that $D^2=V^2+e$, where $e \ge 0$ is the variance of $\psi$. Then, when ~\eqref{eq:stab} is satisfied, the condition~\eqref{eq:stab_H} with $D^2=V^2+e$ is automatically satisfied. In conclusion, the choice of $\psi$ being a Dirac delta is the most unstable one and, therefore, this choice makes computations feasible and allows to predict a wider instability region.

As in the linear case, we may also use the phase $\phi_\varepsilon=-\varepsilon \log \rho_\varepsilon$ in the aggregate equation~\eqref{eq:macro_rho}. Letting $\varepsilon \rightarrow 0 $ and assuming $\phi_\varepsilon \rightarrow \phi$, we formally obtain
\begin{equation}\label{eq:eikonal_adh}
    \partial_t \phi+ \Ub_R^G \cdot \nabla \phi+\textrm{tr}\left[ \mathbb{D}_R^G(\nabla_\x\phi\otimes \nabla_\x \phi)\right]=0,
\end{equation}
because $\Ub_\rho^\varepsilon\to \Ub_R^G(\nabla \phi)$ (defined in~\eqref{eq:URG} and $\mathbb{D}_\rho^\varepsilon \to \mathbb{D}_R^G(\nabla \phi)$, which is the variance covariance matrix of~\eqref{def:G}. Therefore, Eq.~\eqref{eq:eikonal_adh}, in opposition to the linear case, is not the quadratic expansion near $\nabla \phi=0$ of equation~\eqref{HJ_nonlinear} except in the regime $R$ small.



\subsection{The regime $R$ small}

Let us now consider $R$ small in the sense of~\eqref{def:l_S} with $\cS=\rho$, i.e., 
\[
0<R\ll l_\rho := \dfrac{1}{\max \dfrac{|\nabla \rho|}{\rho}},
\]
we may expand $\rho$ as
\[
\rho(t,\x+R\hv) \approx \rho(t,\x)+R\hv\cdot \nabla \rho(t,\x),
\]
so that the normalization function of~\eqref{def:Trho} is
\[
c(t,\x) \approx \rho(t,\x)|\mathbb{S}^{d-1}|.
\]
Therefore Eq.~\eqref{eq:cinetique_adh} becomes
\begin{equation}\label{eq:model_simplified2}
    \dfrac{\partial f}{\partial t}(t,\x,v,\hv) + \vb\cdot \nabla f(t,\x,v,\hv) =  \mu  \, \Big( \rho(t,\x)  (\dfrac{1}{|\mathbb{S}^{d-1}|}+\dfrac{R}{|\mathbb{S}^{d-1}|}\hv \cdot \dfrac{\nabla \rho(t,\x)}{\rho(t,\x)}) - f(t,\x,v,\hv) \Big).
\end{equation}
The latter may be assimilated to a case in which $T[\rho]$ is evaluated in a small perturbation of a nondimensionalized homogeneous configuration set equal to $1$ as $\dfrac{R \nabla \rho}{\rho}$ is small being $R\ll l_\rho$. This is exactly the regime in which the linear stability analysis is performed in \cite{loy2020KRM}.  Considering ~\eqref{eq:model_simplified2} with~\eqref{def:hyp}, we have a linearized kinetic equation. Then we obtain $G_R=1-R\hv\cdot p$ and we recover that $H(p)>0$ is equivalent to the condition $V/\mu R>1$.
\\

In the case of~\eqref{eq:model_simplified2} with~\eqref{def:hyp}, we also remark that 
\begin{equation}\label{eq:macro_adh}
\partial_t \rho^0+\nabla_\x \cdot (\rho^0 \Ub_\rho^0)=0
\end{equation}
and that $\Ub_\rho^0=\Ub_R^G(0)$ and $\mathbb{D}_\rho^0=D_p^2 H(0)$, so that like in the linear case, in this linearized regime obtained for $R$ small, the eikonal equation~\eqref{eq:eikonal_adh} is a quadratic expansion of~\eqref{HJ_nonlinear}.


\subsection{Concentration profile}

Concerning the concentration points dynamics, we remark that, like in the linear case, if $\bar{\x}_i$ is a maximum of $\rho$, then
\begin{equation*}
    \dot{\bar{\x}}_i(t)=\Ub_R^G(\bar{\x}_i(t),p)
\end{equation*}
and, as $p=\nabla_\x \varphi(\bar{\x}_i)=0$, then, as a consequence of~\eqref{eq:nablaH}
\[
\dot{\bar{\x}}(t)=\int_{\mathbb{S}^{d-1}}V_\psi(\hv)\hv \, d\hv,
\]
 that is a nonvanishing quantity in the case where $V_\psi(\hv)$ is not even as a function of $\hv$. In such a case it is possible to observe moving patterns, as showed in~\cite{loy2020KRM}. 
 Again, in the regime of small $R$, that is a linearized case, when the formal aggregate limit is~\eqref{eq:macro_adh} and $\Ub_\rho^0=\Ub_R^G(0)$,  the aggregate limit procedure and the WKB analysis give exactly the same amount of information about the evolution of the maxima.
 \\

\begin{figure}
    \centering
    \includegraphics[width=0.6\textwidth]{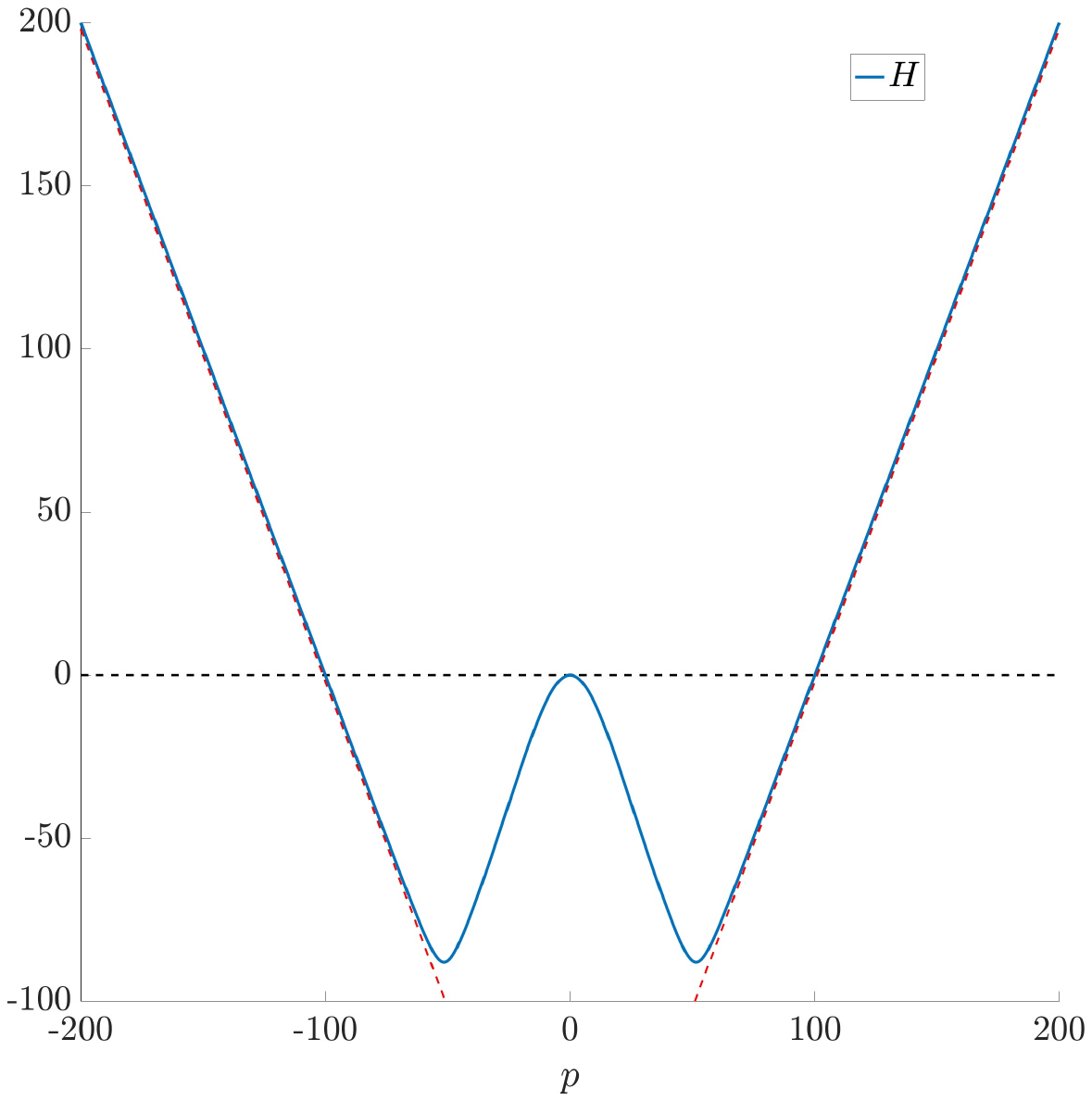}
    \caption{\label{fig:H} The convex-concave Hamiltonian $H$ in the one-dimensional case as computed explicitly in formula~\eqref{eq:H1D}. The values of $\mu, V, R$ are those corresponding to the example in section~\ref{sec:nonlin_ex}. The red dashed lines correspond to the asymptotes $\pm U |p|$, the black horizontal dashed line is the zero-level. The values $\pm \bar p \neq 0$ with $ H(\bar p)=0$ determines the slopes of the saw tooth solutions in Fig.~\ref{fig:adh}}
\end{figure}

 However, the Hamilton-Jacobi equation~\eqref{HJ_nonlinear} also gives the microscopic concentration profile. In the stability regime, i.e. when~\eqref{eq:stab} is satisfied, we actually have that $p=0$ is a minimum of $H$ as $H(0)=0$, $\nabla_p H(0)=0$ when $V_\psi$ is even and $D^2_p(0)>0$. Therefore, we have $H>0$ which implies that $\varphi(t,\x)$ decreases and possible initial concentrations will disappear. In the instability regime, the situation is more interesting and the prototype of the shape of the Hamiltonian is depicted in Fig.~\ref{fig:H}.  In fact, being $D^2_p(0)<0$, then $p=0$ is a maxima. Therefore, there will be a range of values of $p$ where the Hamiltonian is negative. As for $p\rightarrow \pm \infty$, $H \rightarrow \infty$, and if $H$ is $C^1$, there will be a value $\bar p>0$ where $H(\pm \bar p)=0$, and the slopes $\pm \bar p$ determine a saw tooth stationary state since $p=0$ is an unstable state for \eqref{HJ_nonlinear}. This explains the  numerical profile obtained in Fig.~\ref{fig:H}. When $\nabla \varphi$ is small (which is the case near concentration points), then $H(\nabla \varphi) <0$, meaning that $\varphi$ will increase and the concentration will get stronger. 


\commentout{
We know that
\[
f_{\varepsilon}= Q_1(v,\hv,\nabla_\x \varphi)\exp^{-\dfrac{\varphi}{\varepsilon}} \rightarrow \rho_0(t,\x) Q_1(v,\hv,\partial_\x \varphi_0)
\]
We assume that
\[
\rho_\varepsilon \rightharpoonup \sum \omega_i(t) \delta(\x-\x_i(t))=:\rho_0(t,\x)
\]
therefore
\begin{equation}
    \rho_0(t,\x):=\sum \omega_i(t) \delta(\x-\x_i(t))
\end{equation}}

\subsection{An example}\label{sec:nonlin_ex}
As an example, we consider the 1D eigenproblem~\eqref{pb:eigenvalue1} (when $\hv=\pm 1$). Then $H$ is defined by 

\[
1=\dfrac{\mu}{(\exp^{-Rp}+\exp^{Rp})} \int_0^U  \psi(v)\left(\dfrac{\exp^{-Rp}}{\mu+H-vp}+\dfrac{\exp^{Rp}}{\mu+H+vp} \right) \, dv.
\]
Therefore
\[
1=\dfrac{\mu}{2(\exp^{-Rp}+\exp^{Rp})} \int_0^U  \psi(v)\left(\dfrac{\exp^{-Rp}(\mu+H+vp)+\exp^{Rp}(\mu+H-vp)}{(\mu+H)^2-v^2p^2} \right) \, dv .
\]
When $\psi(v|\hv)=\delta(v-\hv V)$, this reduces to
\[
1=\dfrac{\mu}{(\exp^{-Rp}+\exp^{Rp})} \left(\dfrac{\exp^{-Rp}(\mu+H+Vp)+\exp^{Rp}(\mu+H-Vp)}{(\mu+H)^2-V^2p^2} \right) ,
\]
and, therefore
\[
H^2+H\mu-V^2p^2+\mu V p D_R(p)=0, \qquad D_R(p)=\dfrac{\exp^{Rp}-\exp^{-Rp}}{\exp^{-Rp}+\exp^{Rp}}= \tanh(Rp),
\]
in such a way that
\begin{equation}\label{eq:H1D}
H(p)=\dfrac{-\mu+\sqrt{\mu^2+4V^2p^2-4\mu V p D_R(p)}}{2}.
\end{equation}
Consequently, the sign of $H$ is determined by the sign of $V|p|-\mu |D_R(p)|$ and we obtain  
\[
H(p)>0 \qquad \textrm{iff} \qquad \dfrac{V}{\mu}>\dfrac{D_R(Rp)}{p}.
\]
For $|p|$ small this is in accordance with the sign of the second derivative in formula~\eqref{eq:D2H}. 
In the regime when  $R|p|$ is small then $\dfrac{D_R(Rp)}{p}=\dfrac{\tanh(Rp)}{p} \sim R$ and the latter condition becomes~\eqref{eq:stab}. As $p=\nabla \varphi$ and $|\nabla \varphi| \approx \dfrac{|\nabla \rho|}{\rho}$, then it coherently corresponds to the analysis performed in the regime of $R$ small. 
\\

We now show some numerical tests. We solve numerically  the kinetic equation~\eqref{eq:cinetique_adh} in the regime~\eqref{def:hyp}. In particular we choose the following parameter values: $V=1, \mu=100, R=5\cdot 10^{-2}$. Therefore $\varepsilon=10^{-2}$ and we are in the regime of linear instability as $\dfrac{V}{\mu R}=0.2$. We consider three different initial conditions: (a) a perturbation of the homogeneous configuration, (b) a bimodal gaussian (i.e. $\rho^0$ as in~\eqref{S2D}) centered in $\bar{\x}_1=2.3, \bar{\x}_2=2.7$, (c) a bimodal gaussian (i.e. $\rho^0$ as in~\eqref{S2D}) with $\bar{\x}_1=2.4, \bar{\x}_2=2.6$.

\begin{figure}
    \centering
    \hspace{-1.7cm}
    \subfigure[]{\includegraphics[scale=0.22]{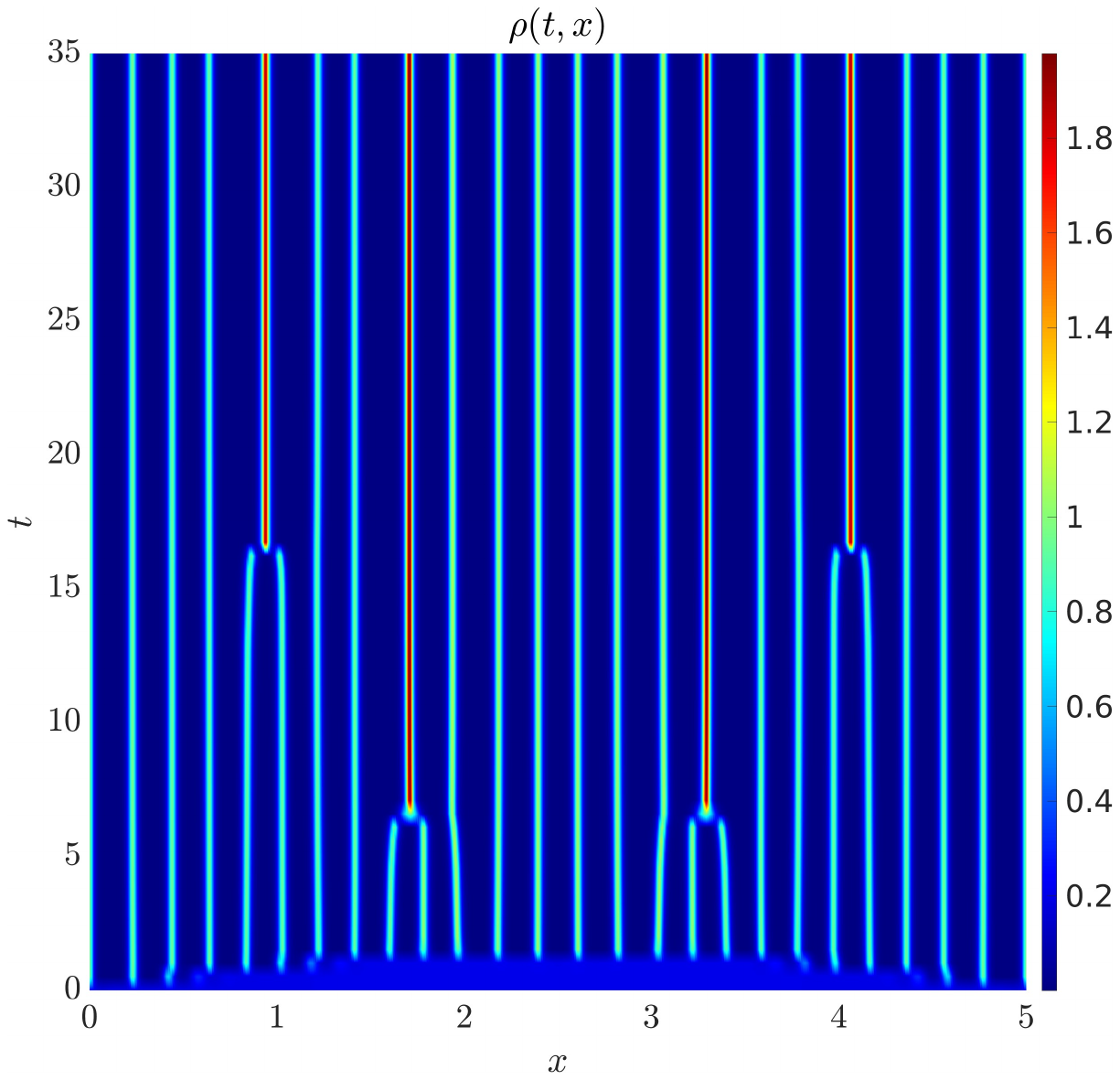}}
    \hspace{-1.8cm}
    \subfigure[]{\includegraphics[scale=0.22]{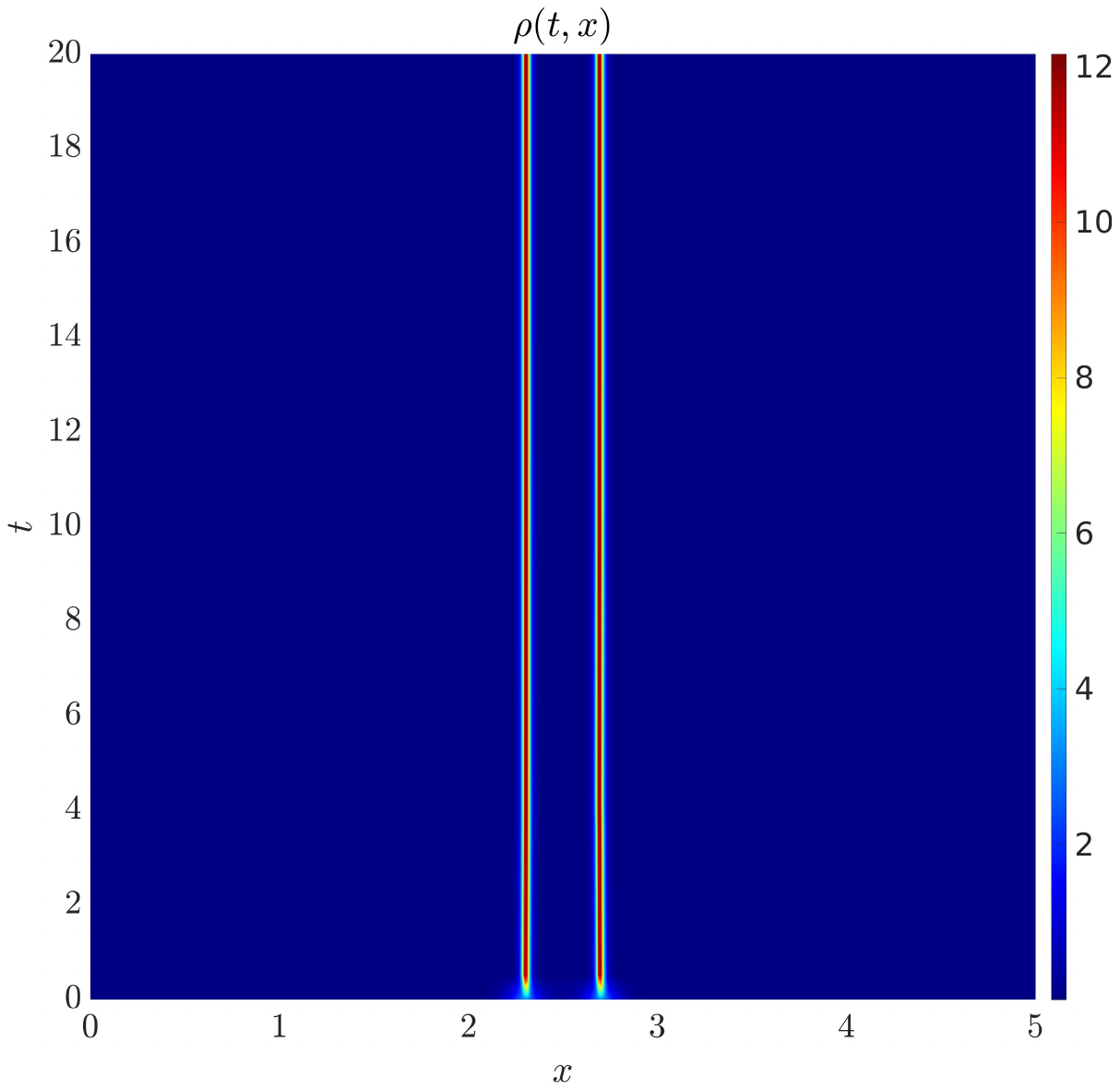}}
    \hspace{-1.8cm}
    \subfigure[]{\includegraphics[scale=0.22]{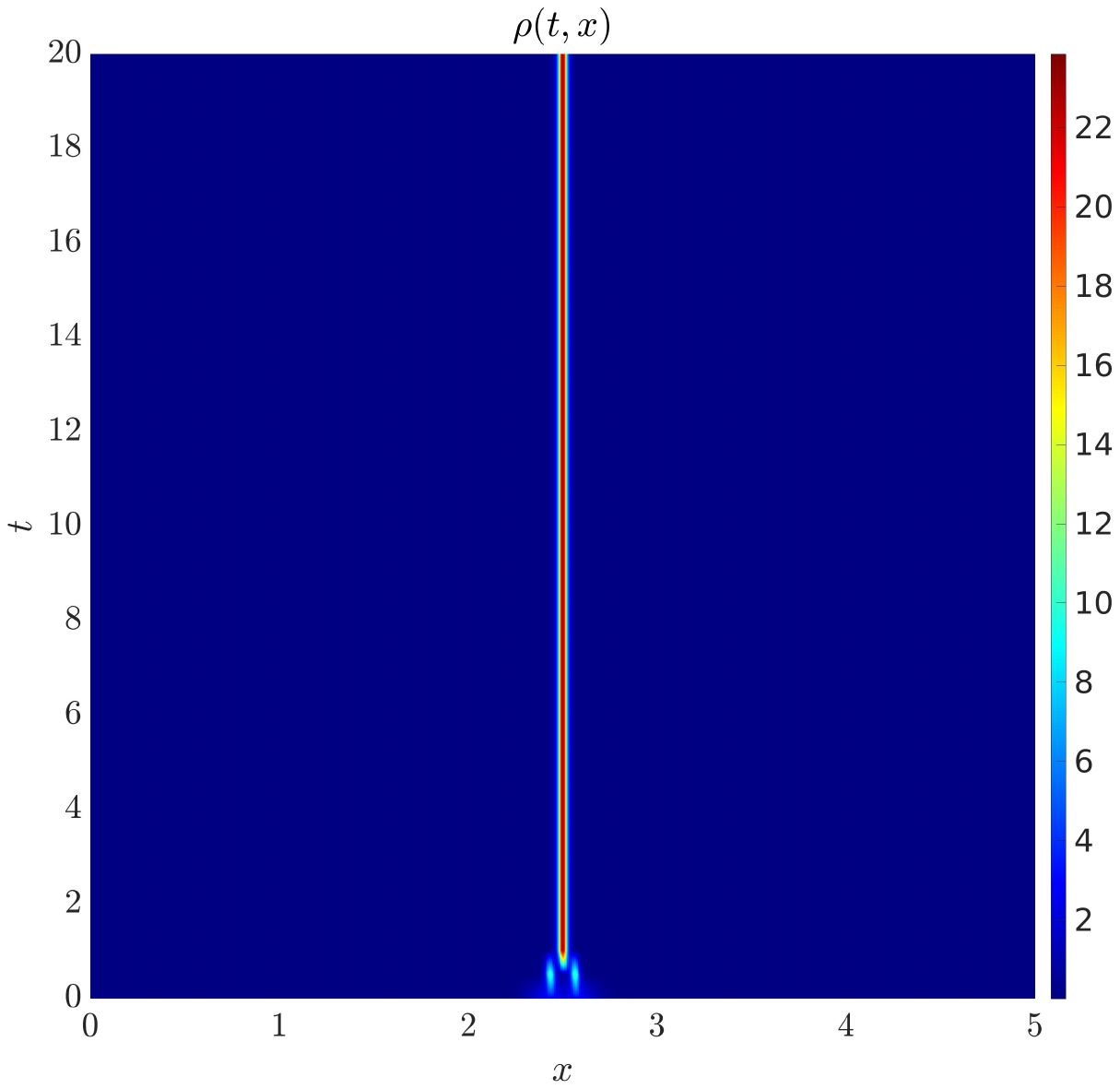}}\\
    \hspace{-1.5cm}
    \subfigure[]{\includegraphics[width=0.3\textwidth]{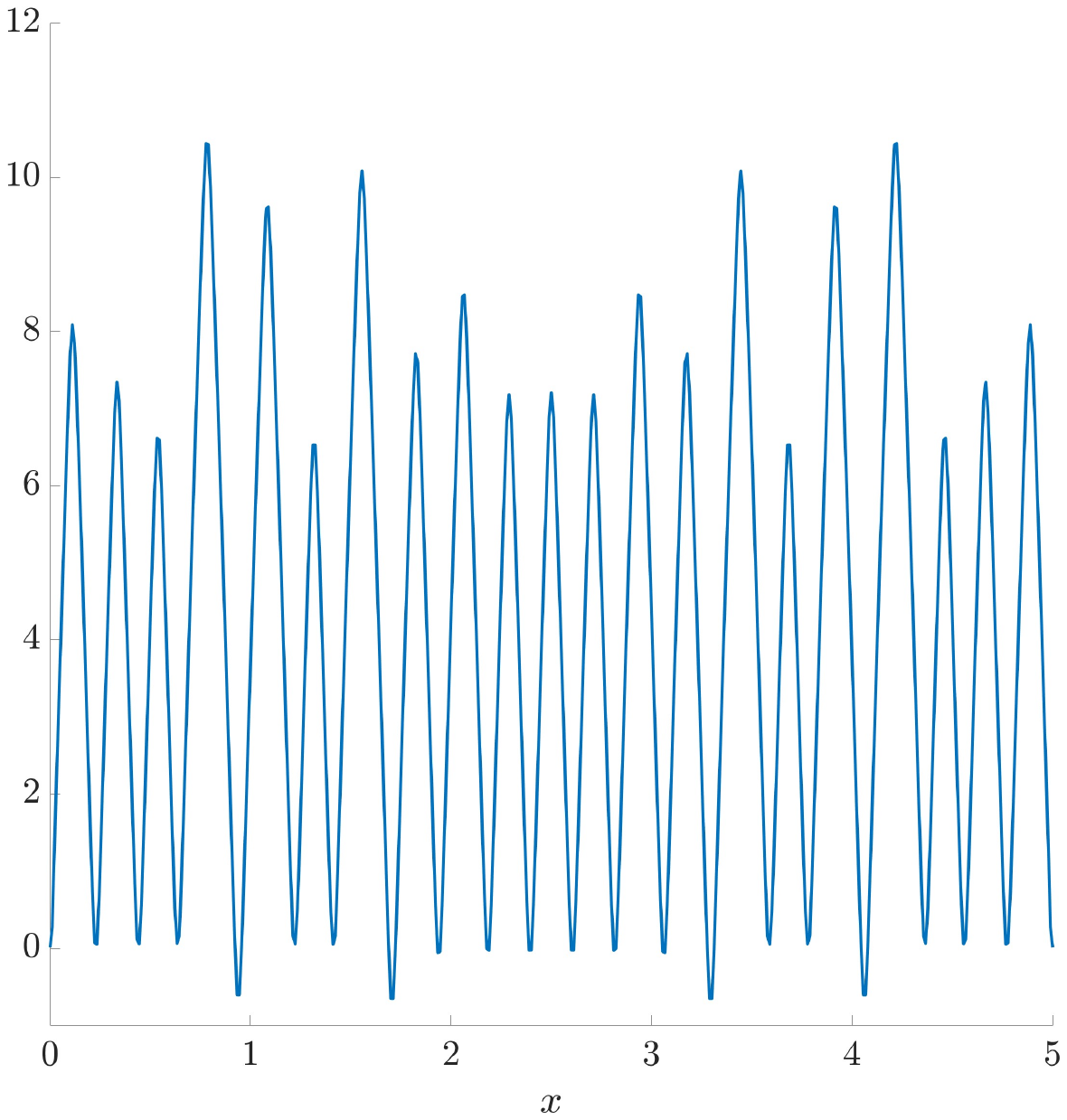}}
    \subfigure[]{\includegraphics[width=0.3\textwidth]{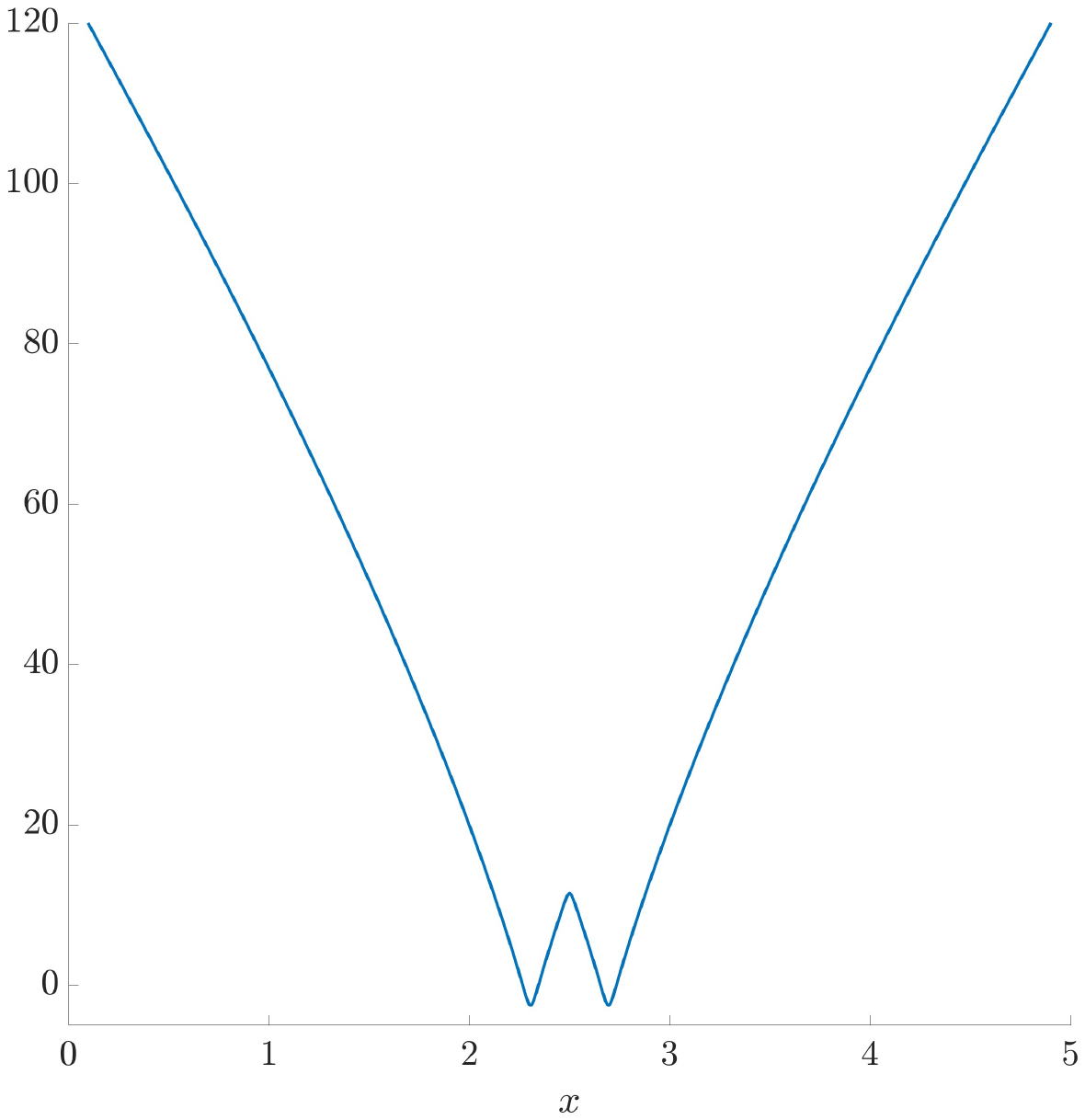}}
    \subfigure[]{\includegraphics[width=0.3\textwidth]{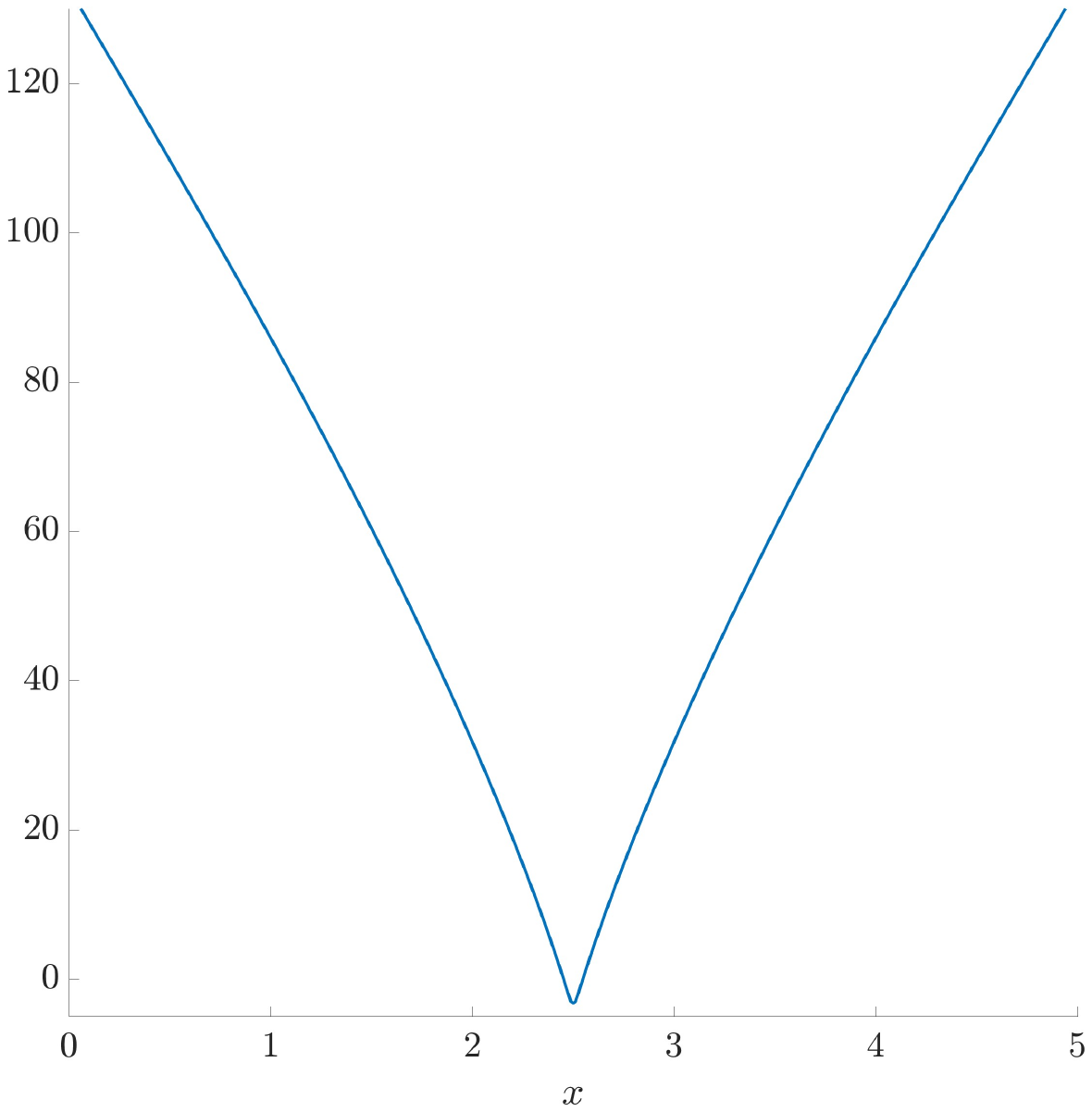}}
    \vspace{-10pt}
    \caption{\label{fig:adh} One dimensional example with parameters $V=1, \mu=100, R=5\cdot 10^{-2}$. In (a) the initial condition is the perturbation of the homogeneous steady profile, in (b) the initial condition is a bimodal gaussian centered in $\bar{\x}_1=2.3, \bar{\x}_2=2.7$, in (c) the initial condition is a bimodal gaussian (i.e. $\rho_0$ as in~\eqref{S2D}) centered in $\bar{\x}_1=2.4, \bar{\x}_2=2.6$. In the second line we report the corresponding profiles of $-\log(\rho(t,\x))$. These saw tooth curves result from the convex-concave Hamiltonian in fig.~\ref{fig:H} }
\end{figure}

As we are in a regime of linear instability, in Fig.\ref{fig:adh}(a) we observe pattern formation, while in figure (b), as the two initial peaks are far enough, they stay so along the dynamics. In Fig.~\ref{fig:adh}(c) we have that the two peaks merge, because the sensing radius is large enough.
In the second line (Fig.~\ref{fig:adh}(d)-(e)-(f), respectively), we plot the corresponding $-\log(\rho)$.

\section{Conclusion}
We have considered a kinetic equation with a BGK relaxation operator in which the transition probability is nonlocal in the physical space which can be linear or nonlinear. For high frequencies, both in the localized and nonlocalized regime, highly concentrated patterns may occur. We analyze them thanks to the WKB ansatz, following~\cite{BC}, and obtain a Hamilton-Jacobi equation in the limit. This method, in the spirit of adaptive dynamics, provides us with the position evolution of the concentration points and with the concentration profile. We can conclude that
\begin{itemize}
    \item In the linear case, the dynamics is driven by an external field $\cS(x)$. At the leading (hyperbolic) order, the kinetic equation and corresponding aggregate limit almost give a complete and coinciding information as ell as the WKB method through the canonical equation for the maxima point. Indeed, we know the position of the maxima; the concentration points do not depend on the initial condition but only on the field $\cS$ as the asymptotic equilibrium is determined by the H theorem and it is independent on the initial condition. The WKB analysis allows also to state the concentration profile around the maxima, 
    in particular when the leading order average velocity of $T$ does not vanish. 

    \item In the nonlinear case, the same link holds between the aggregate limit analysis and the WKB analysis in the linearized regime ($R$ small). When $R$ is not small (and $\rho$ is not smooth) and we expect concentrations, then stating (even formally) the macroscopic limit is not banal. Then, the WKB method gives rise to an unusual convex-concave Hamiltonian, explaining saw tooth patterns which are obtained numerically. It also
    tells us more with respect to the aggregate equation through the canonical equation for the maxima. A difference lies in the fact that it is not possible to find a condition for the location of the concentration points. Moreover, it is possible to recover the linear stability condition found in \cite{loy2020KRM} in the special case of one dimension and $\psi$ a Dirac delta in the limit of $R$ small where $\psi$ being the Dirac delta was chosen in order to determine explicitly the instability condition. Furthermore, the present analysis actually shows that it is possible to extend the results to other distributions $\psi$ that have a larger second moment with respect to the Dirac delta, that is the most unstable one, in the sense that it prescribes a larger parameters region where we have linear instability. Moreover, the analysis can be done in any dimension and the study of the positivity of the Hessian matrix of the Hamiltonian allows to state the same result in any dimension.
\end{itemize}

Both in the linear and nonlinear cases the aggregate and the WKB analysis give compatible eikonal  equations in the suitable limit. However, in the nonlinear case the location of the concentration points cannot be explicitly determined. The WKB analysis, as a further contribution, allows to determine a concentration profile. In the fully nonlinear case ($R$ not small), the WKB analysis can be used in order to understand the dynamics as the analysis of the aggregate limits is not banal and as it goes beyond the regime of the linear stability analysis. In the particular case analysed here we obtain the same results, but we can expect that in other choices of transition probability $T$ the WKB analysis predicts a different region. Moreover, the WKB can be performed only in the \eqref{def:hyp} regime, but $R$ in the dimensional system is not needed to be small like in the linear stability analysis. These considerations suggest that the WKB analysis could be used in order to extend the results of a linear stability analysis to other transition probabilities.


In the context of the study of kinetic eikonal equations in the same spirit as~\cite{BC}, this work has allowed to make some steps further as $i)$ the Markovian probability in the relaxation operator depends on the spatial variable as it is nonlocal in the physical space, $ii)$ it was applied in order to study the space dependent equilibrium in a regime in which concentrations are shown, $iii)$ a regime in which the WKB and aggregate limit procedure may commute was detected.

Moreover, as $T$ depends on $\x$ and on the small parameter $\varepsilon$, in the linear case $H$ depends on both $\x$ and the $\nabla \varphi$, then we obtain a time an evolution equation~\eqref{eq:HJfunc} for $H_\varepsilon$ and, then, a time dependent eigenvalue problem similar to the principal bundle for parabolic equations. Another open problem is to determine the  boundary conditions for the Hamilton-Jacobi equations.
\\

\subsection*{Acknowledgements}
N.L. is member of INdAM-GNFM. N.L. acknowledges support by the Italian Ministry for Education, University and Research (MUR)
through the ``Dipartimenti di Eccellenza" Programme (2018- 2023) of the  Department of Mathematical Sciences, G. L. Lagrange, Politecnico di Torino (CUP: E11G18000350001).
N.L. gratefully acknowledges support from the Italian Ministry of University and Research (MUR) through the grant PRIN2022-PNRR project (No. P2022Z7ZAJ) “A Unitary Mathematical Framework for Modelling Muscular Dystrophies” (CUP: E53D23018070001). 
N.L. gratefully acknowledges support from the CNRS International Research Project ‘Modélisation de la biomécanique cellulaire et tissulaire’ (MOCETIBI).


\section*{Appendix. Boundary conditions} \label{Appendix_cb}

In Section~\ref{sec:aggregate} we have shown the boundary conditions for the hyperbolic limit \eqref{eq:macro_0} of the kinetic equation \eqref{eq:cinetique} that are given by \eqref{eq:macro_cb}. In particular, both \eqref{eq:macro_0} and \eqref{eq:macro_cb} are derived from the kinetic equation: \eqref{eq:macro_0} is derived from \eqref{eq:cinetique_hyp} and \eqref{eq:macro_cb} are derived from \eqref{noflux} that is satisfied by any $f$ that obeys kinetic boundary conditions in the form \eqref{Max_cb}. The kinetic boundary conditions  \eqref{Max_cb} are actually imposed on the entering boundary, i.e. on $\Gamma_-(\x)$. The derived aggregate boundary conditions are noflux boundary conditions for the conservation law \eqref{eq:macro_0}. Actually, we want to verify that those boundary conditions are to be imposed on the entering zone only, i.e. for $\x \in \partial \Omega$ such that $\Ub_\cS^0(\x) \cdot \boldsymbol{n}(\x)<0$ as in the outgoing region, i.e. for $\x \in \partial \Omega$ such that $\Ub_\cS^0 \cdot \boldsymbol{n}>0$ they are granted by the underlying kinetic boundary conditions.

Therefore, we need to compute the average $\Ub_\cS^0(\x)$ for $\x \in \partial \Omega$ that we denote as $\Ub_{\cS_{|\partial \Omega}}^0$. First of all we need to define $T[\cS]_{0_{|\partial \Omega}}$. 
Working in the regime~\eqref{def:hyp_v},
 we define it as
\begin{equation*}
T[\cS]_{0_{|\partial \Omega}} (v,\hv)=c(\x) b(\cS(\x+R(\x,\hv) \hv)) \psi_{|\partial \Omega}(v|\hv), \end{equation*}
where $R(\x,\hv)$ is defined in~\eqref{def:Rxv} and $\psi_{|\partial \Omega}(v|\hv)$ is to be dependent on $\x \in \partial \Omega$, as for $\hv \cdot \boldsymbol{n}(\x)>0$, then we should set $V_\psi=0.$ Therefore, we have that
\begin{equation*}
T[\cS]_{0_{|\partial \Omega}} (v,\hv)=
\begin{cases}
  \dfrac{\psi_{|\partial \Omega}(v|\hv)}{|\mathbb{S}^{d-1}|} \qquad &\textrm{if} \quad \hv \cdot \boldsymbol{n} >0, \\
  c(\x) b(\cS(\x+R\hv)) \psi_{|\partial \Omega}(v|\hv)  \qquad &\textrm{if} \quad \hv \cdot \boldsymbol{n} <0,
\end{cases}
\end{equation*}
as when $\hv \cdot \boldsymbol{n(\x)}>0$ then $R(\x,\hv)=0$, while when $\hv \cdot \boldsymbol{n}<0$, $\x +R \hv \in \Omega$ if, for example $\Omega$ is convex. 

Now, as $T[\cS]_0$ is in fact the equilibrium, it must satisfy the boundary conditions \eqref{Max_cb}. We analyse the two cases $\alpha=0$ (purely Maxwellian) and $\alpha=1$ (pure reflection), any case in between follows as a convex combination.
If we consider the Maxwellian boundary conditions, then we have
\[
M(\x,v,\hv)=c(\x)b(\cS(\x+R(\x,\hv)\hv) \psi(v|\hv), \qquad \hv \in \Gamma_-(\x).
\]
and
\begin{align*}
\Ub_{\cS_{\partial \Omega}}^0 &= \left[  \int_0^U v\left( \int_{\hv \cdot \boldsymbol{n}<0} M(\x,v,\hv) \hv d\hv + \int_{\hv \cdot \boldsymbol{n}>0}\dfrac{1}{|\mathbb{S}^{d-1}|} \psi(v|\hv) \hv d\hv \right) dv \right]\\
&=  \int_0^U v \int_{\hv \cdot \boldsymbol{n}<0} M(\x,v,\hv) \hv d\hv dv. 
\end{align*}
Therefore
\[
\Ub_{\cS_{\partial \Omega}}^0 \cdot \boldsymbol{n} =\int_0^U v \int_{\hv \cdot \boldsymbol{n}<0} M(\x,v,\hv) \hv \cdot \boldsymbol{n} d\hv dv <0. 
\]
In conclusion the whole boundary is an entering zone and then we need to impose~\eqref{eq:cb_macro}.
If $\alpha=1$, then $T[\cS]_0$ must satisfy the following boundary conditions, if $\hv\in \Gamma_-(\x)$
\[
T[\cS](v,\hv)_{\Gamma_-}=T[\cS](v,\mathcal{W}(\hv)),
\]
and here $\mathcal{W}(\hv) \cdot \boldsymbol{n} >0$. Therefore $T[\cS]_{|\Gamma_-}=\dfrac{\psi(v|\hv)}{|\mathbb{S}^{d-1}|}$. In conclusion, following the same computations as for the case $\alpha=0$, we find $\Ub_\cS^0=0,$ i.e. the velocity vector vanishes on the whole boundary and \eqref{eq:cb_macro} is satisfied. 

Conversely, in the regime \eqref{def:hyp}, we have that if $T[\cS]_0$
does not depend on $\hv$ (and on \x) because of the localization, then $\Ub_\cS^0=0$ on $\partial \Omega$.




\bibliography{references}

\begin{thebibliography}{10}

\bibitem{Armstrong_Painter_Sherratt.06}
N.~J. Armstrong, K.~J. Painter, and J.~A. Sherratt.
\newblock A continuum approach to modelling cell-cell adhesion.
\newblock {\em Journal of theoretical biology}, 243 1:98--113, 2006.

\bibitem{GB:94}
G.~Barles.
\newblock {\em {Solutions de viscosit{\'e} des {\'e}quations de
  Hamilton-Jacobi}}.
\newblock Springer-Verlag Berlin Heidelberg, 1994.

\bibitem{BaPe2015}
G.~Barles and B.~Perthame.
\newblock {Concentrations and constrained Hamilton-Jacobi equations arising in
  adpative dynamics}.
\newblock {\em Contemporary Mathematics}, 439:57, 2007.

\bibitem{bouin2015KRM}
E.~Bouin.
\newblock A hamilton-jacobi approach for front propagation in kinetic
  equations, 2015.

\bibitem{bouin2019EJAM}
E.~Bouin and N.~Caillerie.
\newblock Spreading in kinetic reaction–transport equations in higher
  velocity dimensions.
\newblock {\em European Journal of Applied Mathematics}, 30(2):219–247, 2019.

\bibitem{BC}
E.~Bouin and V.~Calvez.
\newblock A kinetic eikonal equation.
\newblock {\em Comptes Rendus Mathematique}, 350(5):243--248, 2012.

\bibitem{bouin2023JLMS}
E.~Bouin, V.~Calvez, E.~Grenier, and G.~Nadin.
\newblock Large-scale asymptotics of velocity-jump processes and nonlocal
  hamilton–jacobi equations.
\newblock {\em Journal of the London Mathematical Society}, 108(1):141--189,
  2023.

\bibitem{bouin2015ARMA}
E.~Bouin, V.~Calvez, and G.~Nadin.
\newblock Propagation in a kinetic reaction-transport equation: Travelling
  waves and accelerating fronts.
\newblock {\em Archive for Rational Mechanics and Analysis}, 217, 08 2015.

\bibitem{Caillerie}
N.~Caillerie.
\newblock Large deviations of a forced velocity-jump process with a
  {Hamilton{\textendash}Jacobi} approach.
\newblock {\em Annales de l'Institut Fourier}, 71(4):1733--1755, 2021.

\bibitem{Calvez2015KRM}
V.~Calvez, G.~Raoul, and C.~Schmeiser.
\newblock Confinement by biased velocity jumps: Aggregation of escherichia
  coli.
\newblock {\em Kinetic and Related Models}, 8(4):651--666, 2015.

\bibitem{Cercignani}
C.~Cercignani.
\newblock {\em The Boltzmann Equation and its Applications}.
\newblock Springer, New York, 1987.

\bibitem{Chalub_Markowich_Perthame_Schmeiser.04}
F.~A. C.~C. Chalub, P.~A. Markowich, B.~Perthame, and C.~Schmeiser.
\newblock Kinetic models for chemotaxis and their drift-diffusion limits.
\newblock {\em Monatshefte f{\"u}r Mathematik}, 142(1):123--141, Jun 2004.

\bibitem{Chauviere_Hillen_Preziosi.07}
A.~Chauviere, T.~Hillen, and L.~Preziosi.
\newblock Modeling cell movement in anisotropic and heterogeneous network
  tissues.
\newblock {\em Networks $\&$ Heterogeneous Media}, 2(2):333, 2007.

\bibitem{Chen2020PTRSB}
L.~Chen, K.~Painter, C.~Surulescu, and A.~Zhigun.
\newblock Mathematical models for cell migration: a non-local perspective.
\newblock {\em Philosophical Transactions of the Royal Society B: Biological
  Sciences}, 375(1807):20190379, 2020.

\bibitem{conte2023SIAP}
M.~Conte and N.~Loy.
\newblock A non-local kinetic model for cell migration: A study of the
  interplay between contact guidance and steric hindrance.
\newblock {\em SIAM Journal on Applied Mathematics}, 0(0):S429--S451, 0.

\bibitem{conte2022BMB}
M.~Conte and N.~Loy.
\newblock Multi-cue kinetic model with non-local sensing for cell migration on
  a fiber network with chemotaxis.
\newblock {\em Bull Math Biol.}, 84(3), 2022.

\bibitem{crandall1992}
M.~G. Crandall, H.~Ishii, and P.-L. Lions.
\newblock {User's guide to viscosity solutions of second order partial
  differential equations}.
\newblock {\em Bulletin of the American Mathematical Society}, 27(1):1--67,
  1992.

\bibitem{evans1989}
L.~Evans.
\newblock The perturbed test function method for viscosity solutions of
  nonlinear {PDE}.
\newblock {\em Proc. Roy. Soc. Edinburgh Sect. A}, 111(3-4):359--375, 1989.

\bibitem{Filbet_Perthame}
F.~Filbet, P.~Laurencot, and B.~Perthame.
\newblock Derivation of hyperbolic models for chemosensitive movement.
\newblock {\em Journal of Mathematical Biology}, 50:189--207, 03 2005.

\bibitem{Filbet}
F.~Filbet and N.~Vauchelet.
\newblock Numerical simulation of a kinetic model for chemotaxis.
\newblock {\em Kinetic and Related Models}, 3:B348--B366, 09 2010.

\bibitem{Hillen.05}
T.~Hillen.
\newblock M5 mesoscopic and macroscopic models for mesenchymal motion.
\newblock {\em Journal of mathematical biology}, 53:585--616, 11 2006.

\bibitem{Othmer_Hillen.00}
T.~Hillen and H.~G. Othmer.
\newblock The diffusion limit of transport equations derived from velocity-jump
  processes.
\newblock {\em SIAM Journal of Applied Mathematics}, 61:751--775, 2000.

\bibitem{hillen2007DCDSB}
T.~Hillen, K.~Painter, and C.~Schmeiser.
\newblock Global existence for chemotaxis with finite sampling radius, 2007.

\bibitem{Keller_Segel}
E.~F. Keller and L.~A. Segel.
\newblock Initiation of slime mold aggregation viewed as an instability.
\newblock {\em Journal of Theoretical Biology}, 26(3):399 -- 415, 1970.

\bibitem{LamLou}
K.~Lam and Y.~Lou.
\newblock {\em Introduction to Reaction-Diffusion Equations}.
\newblock Lecture Notes on Mathematical Modelling in the Life Sciences.
  Springer International Publishing, 2022.

\bibitem{LLP2023}
K.~Lam, Y.~Lou, and B.~Perthame.
\newblock A {H}amilton-{J}acobi approach to evolution of dispersal.
\newblock {\em Communications in Partial Differential Equations},
  48(1):86--118, 2023.

\bibitem{Lods}
B.~Lods.
\newblock Semigroup generation properties of streaming operators with
  noncontractive boundary conditions.
\newblock {\em Mathematical and Computer Modelling}, 42:1441--1462, 12 2005.

\bibitem{Lorenzi_2020}
Tommaso Lorenzi and Camille Pouchol.
\newblock Asymptotic analysis of selection-mutation models in the presence of
  multiple fitness peaks.
\newblock {\em Nonlinearity}, 33(11):5791, oct 2020.

\bibitem{AL_SM_BP}
A.~Lorz, S.~Mirrahimi, and B.~Perthame.
\newblock {Dirac mass dynamics in multidimensional nonlocal parabolic
  equations}.
\newblock {\em Communications in Partial Differential Equations},
  36(6):1071--1098, 2011.

\bibitem{loy2021EJAM}
N.~Loy, T.~Hillen, and K.~Painter.
\newblock Direction dependent turning leads to anisotropic diffusion and
  persistence.
\newblock {\em European Journal of Applied Mathematics}, 33(4):729--765, 2022.

\bibitem{loy2019JMB}
N.~Loy and L.~Preziosi.
\newblock Kinetic models with non-local sensing determining cell polarization
  and speed according to independent cues.
\newblock {\em Journal of Mathematical Biology}, 80:373--421, 2020.

\bibitem{loy2020JMB}
N.~Loy and L.~Preziosi.
\newblock Modelling physical limits of migration by a kinetic model with
  non-local sensing.
\newblock {\em Journal of Mathematical Biology}, 80, 2020.

\bibitem{loy2020KRM}
N.~Loy and L.~Preziosi.
\newblock Stability of a non-local kinetic model for cell migration with
  density dependent orientation bias.
\newblock {\em Kinetic and Related Models}, 13(5):1007--1027, 2020.

\bibitem{Othmer_Hillen.02}
H.~Othmer and T.~Hillen.
\newblock The diffusion limit of transport equations ii: Chemotaxis equations.
\newblock {\em SIAM Journal of Applied Mathematics}, 62:1222--1250, 04 2002.

\bibitem{Alt.88}
H.~G. Othmer, S.~R. Dunbar, and W.~Alt.
\newblock Models of dispersal in biological systems.
\newblock {\em Journal of Mathematical Biology}, 26(3):263--298, Jun 1988.

\bibitem{Plaza}
R.G. Plaza.
\newblock Derivation of a bacterial nutrient-taxis system with doubly
  degenerate cross-diffusion as the parabolic limit of a velocity-jump process.
\newblock {\em Journal of mathematical biology}, 2019.

\bibitem{Stroock}
D.~W. Stroock.
\newblock Some stochastic processes which arise from a model of the motion of a
  bacterium.
\newblock {\em Zeitschrift f{\"u}r Wahrscheinlichkeitstheorie und Verwandte
  Gebiete}, 28(4):305--315, Dec 1974.

\end{thebibliography}
\end{document}